\documentclass[twocolumn, abstract]{aa} %
\usepackage{graphicx}
\usepackage{natbib}
\bibpunct{(}{)}{;}{a}{}{,}
\usepackage{syntonly}
\usepackage[fleqn]{mathtools}
\usepackage{amssymb}
\usepackage{amsmath}
\usepackage{url}

\usepackage{layouts}
\usepackage{color}
    \definecolor{Blue}{rgb}{0.0,0.0,1.0}
    \definecolor{Red}{rgb}{1.0,0.0,0.0}
    \definecolor{Green}{rgb}{0.0,1.0,0.0}
\newcommand\red{\textcolor[rgb]{0.0,0.0,0.0}}
\newcommand\blue{\textcolor[rgb]{0.0,0.0,0.0}}

\newcommand\myred{\textcolor[rgb]{0.0,0.0,0.0}}

\newcommand{\be}{\begin{equation}}
\newcommand{\ee}{\end{equation}}
\newcommand{\bea}{\begin{eqnarray}}
\newcommand{\eea}{\end{eqnarray}}
\newcommand{\nn}{\nonumber}
\newcommand{\pp}{\varphi}
\newcommand{\dd}{\mathrm{d}}
\renewcommand\thesection{\arabic{section}}
\begin{document}
%-----------------------------------------------------------------------
\title{X-ray-binary spectra in the lamp post model}

%
%\subtitle{no subtitle yet}
%-----------------------------------------------------------------------
\author{     F. H. Vincent\inst{1,2}
%-----------------------------------------------------------------------
\and          A. R\'o\.{z}a\'nska\inst{2}
%-----------------------------------------------------------------------
\and		A. A. Zdziarski\inst{2}
%-----------------------------------------------------------------------
\and		J. Madej\inst{3}
}
%-----------------------------------------------------------------------
%-----------------------------------------------------------------------
\institute{
	LESIA, Observatoire de Paris, CNRS, Universit\'e Pierre et Marie Curie, Universit\'e Paris Diderot, 5 place Jules Janssen,
			92190 Meudon, France
	 \\ \email{frederic.vincent@obspm.fr}
%-----------------------------------------------------------------------
\and
              Nicolaus Copernicus Astronomical Center, ul. Bartycka 18, PL-00-716 Warszawa, Poland
\and
         	Astronomical Observatory, University of Warsaw, Al. Ujazdowskie 4, 00-478 Warsaw, Poland
}
%-----------------------------------------------------------------------
   \date{Received ; accepted }
%-----------------------------------------------------------------------
%-----------------------------------------------------------------------
\abstract
%............................................................C O N T E X T
   {The high-energy radiation from black-hole binaries may be due to the reprocessing of a \textit{lamp}
   located on the black hole rotation axis and emitting X-rays.
   The observed spectrum is made of three major components: the \textit{direct} spectrum traveling from
   the lamp directly to the observer; the 
   \textit{thermal bump} at the equilibrium temperature of the accretion disk heated by the lamp; and the 
   \textit{reflected spectrum} essentially made of the Compton hump and the iron-line complex.
   }
%..................................................................A I M S
   {We aim at computing accurately the complete reprocessed spectrum (thermal bump + reflected) of black-hole binaries over the entire X-ray band.
   \red{We also determine the strength of the direct component.}
   \red{Our choice of parameters is adapted to a source showing an important thermal component.}
   We are particularly interested in investigating the possibility to use the iron-line complex as a probe to constrain the
   black hole spin.}
%............................................................M E T H O D S
   {We compute in full general relativity the illumination of a thin accretion disk by a fixed X-ray lamp along the rotation axis.
   We use the ATM21 radiative transfer code to compute the local, energy-dependent spectrum emitted along the disk as a function of radius, emission angle and black hole spin. We then ray trace this local spectrum to determine the final \red{reprocessed} spectrum as \red{received} by a distant observer. We consider two extreme values of the black hole spin ($a=0$ and $a=0.98$) and discuss the dependence of the local and ray-traced spectra
   on the emission angle and black hole spin.
   }
%............................................................R E S U L T S
   {We show the importance of the angle dependence of the total disk \blue{specific intensity} spectrum emitted by the illuminated atmosphere
   when the thermal disk emission if fully taken into account. The disk flux, together with 
   the X-ray flux from the lamp, determines the temperature and ionization structure of 
   the atmosphere. High black hole spin implies high temperature in the inner disk regions, therefore 
   the emitted thermal disk spectrum fully covers the iron-line complex. As a result, instead
   of fluorescent iron emission line, we locally observe absorption lines produced in the hot disk atmosphere. Absorption lines are narrow and disappear 
   after ray tracing the local spectrum.}
%....................................................C O N C L U S I O N S
   {Our results mainly highlight the importance of considering the angle dependence
   of the local spectrum when computing \red{reprocessed} spectra, as was already found in a recent study.
   The main new result of our work is to show the importance of computing the thermal bump
   of the spectrum, as this feature can change considerably the observed iron-line complex. Thus, in particular
   for fitting black hole spins, the full spectrum, and not only the reflected part, should be computed self-consistently.}
%-----------------------------------------------------------------------
%-----------------------------------------------------------------------
%-----------------------------------------------------------------------
%-----------------------------------------------------------------------
%-----------------------------------------------------------------------
%-----------------------------------------------------------------------
%\titlerunning{A magnetized torus for modeling Sgr~A*}
\maketitle
%-----------------------------------------------------------------------
%-----------------------------------------------------------------------
%
%
\section{Introduction}

It is probable that X-ray spectra emitted by black-hole binaries and active galactic nuclei 
contain relativistic signatures due to the proximity of the emission region to the central
black hole. In particular, the distortion of the Fe~K line has been since a long
time advocated to be due to relativistic effects~\citep{fabian89,reynolds03}. 
This feature of the spectrum is used to constrain the spin
parameter of black holes~\citep{reynolds14}.

The observed X-ray spectra show two components, a soft
and a hard one, that can be naturally explained in the scenario
where an accretion disk is accompanied by some Comptonizing region,
which can be typically either a hot corona sandwiching the inner region of the disk~\citep{haardt91},
a hot inner flow occupying the regions close to the black hole with the disk being
truncated to some radius~\citep{zdziarski04,done07},
or the base of a jet located on the rotation axis of the black hole~\citep[see among many others][]{matt91,martocchia96,markoff05}. 
In this article
we will only consider this third option, the so-called lamp post model.

Our aims are (1) to compute the flux of illuminating radiation from this lamp
to the accretion disk, 
(2) to compute the reflected \blue{intensity} spectrum due to the
reprocessing of the illuminating flux by the disk and (3) to ray trace this to a
distant observer. 
%\red{We note that we are not considering the direct component
%of the spectrum, which reaches the observer without any reprocessing by the disk.}
Many works have been investigating these various points in the past.
Point (1) and (3) essentially deal with the effect of light bending from a lamp post 
to the illuminated accretion disk and then to the distant 
observer~\citep[see among others][]{reynolds99,ruszowski00,miniutti03,miniutti04,fukumura07,wilkins12,dauser13,dovciak14}.
Point (2) requires solving the complex radiative transfer in the illuminated
accretion disk and has also been investigated by many authors since quite
a long time, taking into account either Newtonian or general-relativistic 
illumination, solving the radiative transfer equations in a Newtonian 
framework, and assuming the reflection from a constant-density slab~\citep[see among others][]{george91,matt91,ross93,zycki94,czerny94,magdziarz95,poutanen96,niedzwiecki08,garcia10}. 
In addition, the assumption of a stratified gas being in hydrostatic equilibrium is more 
realistic but more complex from a numerical point of view 
\citep{nayakshin01,ballantyne01,rozanska02,rozanska08,rozanska11}.
\blue{In particular, \citet[][together with the subsequent works of the same authors]{rozanska08} solve the
radiative equilibrium which was not dealt with previously, to our knowledge, by any other group.}
%{\it \bf I do not see any refs for point (3) to your papers or Dovciak or others }.
{The most advanced attempt to take into account points (1), (2)
and (3) is the work of~\citet{garcia14}}, %Taking into account self-consistently points (1), (2)
%and (3) was so far only done by~\citet{garcia14}, 
which was the main motivation for the
development of this article. These authors, however, still consider the reflection 
from a constant-density slab and that the
illumination is incident on the disk at a constant angle equal to $45^\circ$. 

This article's treatment of point (1) and (3) is {close} to the treatment
of~\citet{dauser13} and~\citet{garcia14}. {However, there are two important
differences. The first one is that we take into account the redshift effect on the
frequency cut-offs assumed in the lamp frame~\citep[see the discussion in][]{niedzwiecki16}.
The second one is that we do not assume a $45^\circ$ incidence of the illumination but
rather consider the true direction of incidence to compute the mean illuminating intensity.}
As far as point (2) is concerned, there is an important difference between our treatment
and the one of~\citet{garcia14}, which is the main reason for us to revisit this topic.
We are computing local \blue{intensity} spectra 
using the code ATM21 of~\citet{rozanska11}. This code solves the hydrostatic \blue{and radiative} equilibria
of the disk self-consistently, taking into account a slim disk solution~\citep{sadowski11} 
with a given spin parameter. The full spectrum, including the thermal bump due to
the radiation of the accretion disk heated by the lamp, is computed.
\citet{garcia14} do not solve the hydrostatic
equilibrium and rather assume a constant-density slab. 
Moreover, they only compute the reflected part of the spectrum
and do not consider the thermal bump.
As far as {the value of} spin is concerned, this means that only the inner
radius of the disk (supposed to be at the innermost stable circular orbit, the ISCO, which
depends on the spin parameter) takes spin into account in~\citet{garcia14}. Besides this,
the disk physics is the same whatever the spin.
This is an important simplification
and it was highlighted by \citet{nayakshin01} that the two options 
(considering a constant-density slab or solving the hydrostatics of the disk) do not lead to the
same predictions. Moreover, as we are particularly interested in taking as much as
possible general-relativistic effects into account (particularly spin effects), 
we believe that it is important to solve
the hydrostatic equilibrium for the particular value of spin parameter assumed for the central
black hole. We note that our treatment is not {100\%} relativistic as the radiative transfer
is solved in a Newtonian spacetime~\citep[similarly as in][]{garcia14}. However, given that point (2) is restricted to local 
phenomena, we do not think that this may have important consequences on the modeled spectra.
%{\bf This treatmant of Newtonian local radiative transfer computation is also used in  
 %\citet{garcia14}  paper}. 
We thus believe that the simulations we
present are among the most realistic to date, as far as taking spin effect into account is concerned.
Understanding spin effects imprinted in the reflected spectrum is particularly important
given that the Fe K line is one of the probe of black hole spins and will be
used by high-precision future mission like {\it \red{Hitomi}}~\citep[\red{formerly ASTRO-H},][]{takahashi14,miller14,reynolds14b} and {\it ATHENA}~\citep{nandra13,dovciak13}.

This article aims at simulating accurately the \red{observed} spectrum generated by a
simple lamp post scenario. Our first goal is to compute accurately the flux emanating from 
the lamp and illuminating the disk (Section~\ref{sec:illum}).
We will compute the illuminating flux either in a Newtonian spacetime or in the Kerr spacetime
with spin parameter $a=0$ and $a=0.98$. This allows us to compare the resulting \blue{local specific intensity reflected
spectrum, emitted by the disk at some emission angle towards the Earth}, for these three kinds of illuminations.
Our second goal is to compute the local reflected spectrum due to the reprocessing
of the illuminating flux by the disk (Section~\ref{sec:localspec}). Our third goal is to ray trace this local spectrum to
a distant observer (Section~\ref{sec:raytrace}).
Finally, Section~\ref{sec:conclu} gives conclusions and perspectives.

\section{Illuminating mean intensity}
\label{sec:illum}

We consider a spherical source of radiation (the lamp) located along the rotation axis of a black hole, at some coordinate $z$.
The radius of the spherical source is assumed small enough so that it is perceived
as point-like by any observer rotating with the disk. In our simulations, this radius is $R = 0.25\,M$,
where $M$ is the black hole mass.
We consider {a set of corotating} observers\footnote{Note that these observers are different from the
final observer, located on Earth at a very large distance away from the black hole.} in the disk at several coordinate radii $r$.
The global accretion disk model considered for computing the reflected spectrum is a slim disk 
as described by~\citet{sadowski11} with an accretion rate $\dot{m} = 0.01\,\dot{m}_{\mathrm{Edd}}$, where the Eddington accretion rate defined as 
$\dot{m}_{\mathrm{Edd}} = 16\,L_{\mathrm{Edd}}/c^2$. For this \red{moderate} accretion
rate, the disk is very thin ($H/r \approx 0.05$) and the gas follows nearly Keplerian orbits. We thus consider
a set of Keplerian observers, corotating with the disk at the local Keplerian angular velocity.
For each of these observers we perform backwards-in-time ray tracing
from the observer towards the lamp in order to determine the illuminating flux received by each observer.

For all ray-tracing computations presented in this article, we use 
the open-source\footnote{Freely available at \url{http://gyoto.obspm.fr}} 
GYOTO code~\citep{vincent11}. 
Photons are traced
by integrating the geodesic equation
using a Runge-Kutta-Fehlberg adaptive-step
integrator at order 7/8 (meaning that the method is 8th order, with an
error estimation at 7th order).

\begin{figure*}[htbp]
\centering
\includegraphics[width=\textwidth,height=0.35\textwidth]{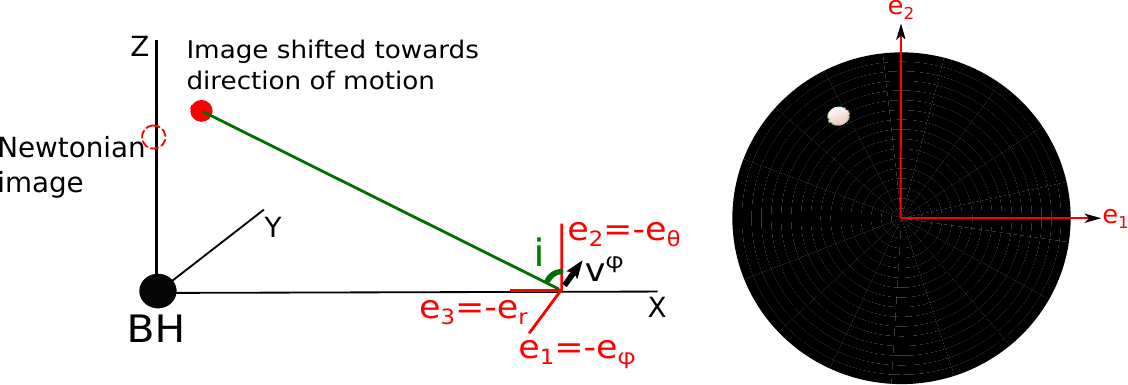}
\caption{\textbf{Left:} geometry of the lamp post model. The velocity $v^{\pp}$ is the Keplerian
velocity of the corotating observer. The triad ($\boldsymbol{e_1}$,$\boldsymbol{e_2}$,$\boldsymbol{e_3}$)
is the local rest frame of the observer. The angle $i$ lies between the direction of the local normal and the
direction of the lamp as seen by the observer. \textbf{Right:} image (map of the quantity $g^{3+\alpha}$ as defined
in Eq.~\ref{eq:Iobs}) of the lamp as
observed by a Keplerian observer rotating at $r=40 \, M$. The total field of view is of $0.9$~rad. Note that the lamp
is not strictly speaking point-like so the angle $i$ is equal to the average of all the directions on sky connecting
the observer to the lamp. Note also that the lamp appears shifted with respect to the local normal because of the
special relativistic effect linked to the high velocity of the observer. The displacement is in the direction of the observer's
motion.}
\label{fig:geom}
\end{figure*}

\myred{We consider a lamp emitting isotropically in its rest frame. This is obviously a simplifying hypothesis
that is most likely not to be satisfied in a realistic context. Indeed, the X-ray radiation from the lamp is likely
to be produced by the Comptonization of photons emitted by the accretion disk~\citep[see e.g. the recent discussion in][]{dovciak15}. This process would lead to
anisotropic radiation. Moreover, we also neglect the motion of the source and it has already been discussed
in the literature that the lamp motion affects the reflected spectrum~\citep{wilkins12,dauser13}. A realistic treatment of the Comptonization
of the disk's photons by a moving source goes beyond the scope of the present paper.}
The lamp is thus emitting radiation isotropically in its rest frame following a power-law spectrum
over the X-ray band
\be
I_\nu^{\mathrm{lamp}} = A \nu_{\mathrm{lamp}}^{-\alpha}, \quad \nu_{\mathrm{lamp}} \in [\nu_{\mathrm{lamp,1}} , \nu_{\mathrm{lamp,2}} ]
\ee
where $\nu_{\mathrm{lamp}}$ is the emitted frequency as measured in the rest frame
of the lamp, $\alpha$ is the spectral index and $A$ is a normalizing constant to
be determined later on. The frequencies $\nu_{\mathrm{lamp,1}}$ and $\nu_{\mathrm{lamp,2}} $
are the limits of the illumination band.
%Unless otherwise stated, the spectral index is fixed to $\alpha=0.7$
%in the whole article.
Once the ray-tracing computation has been performed,
it is possible to determine what is the angle $i$ in each Keplerian observer's rest frame
between the local normal and the direction of the lamp on sky, as well as the value
of the specific intensity received by each Keplerian observer corotating at coordinate radius $r$
\be
\label{eq:Iobs}
I_\nu^{\mathrm{disk}} (i,r) = A g^{3+\alpha} \nu_{\mathrm{disk}}^{-\alpha}, \quad \nu_{\mathrm{disk}} \in [\nu_{\mathrm{disk,1}} , \nu_{\mathrm{disk,2}} ]
\ee
where $\nu_{\mathrm{disk}}$ is the observed frequency, as measured in the rest frame
of the Keplerian observer and $g = \nu_{\mathrm{disk}} / \nu_{\mathrm{lamp}}$ is the redshift factor.
In particular, the limiting frequencies in the disk frame are obtained from their lamp-frame
counterparts by using $\nu_{\mathrm{disk,i}} = g \,\nu_{\mathrm{lamp,i}}$. 
Figure~\ref{fig:geom} illustrates the geometry and the image of the lamp as seen 
by the given Keplerian observer.

The input quantity of the code we use to compute reflected spectra is the mean intensity
\bea
\label{eq:meanintens}
\mathcal{I}^{\mathrm{disk}}_\nu(r) &=&  \frac{1}{4\pi}\int I^{\mathrm{disk}}_\nu \,\dd \Omega \\ \nn
&=&  {A} \nu_{\mathrm{disk}}^{-\alpha} \times \frac{1}{4\pi}\int  g^{3+\alpha} \,\dd \Omega \\ \nn
\eea
where the integration is performed over the local solid angle covered by the lamp (note that
the lamp is not strictly point-like)
as measured in the observer's rest frame.

\begin{figure*}[htbp]
\centering
\includegraphics[width=6cm,height=6cm]{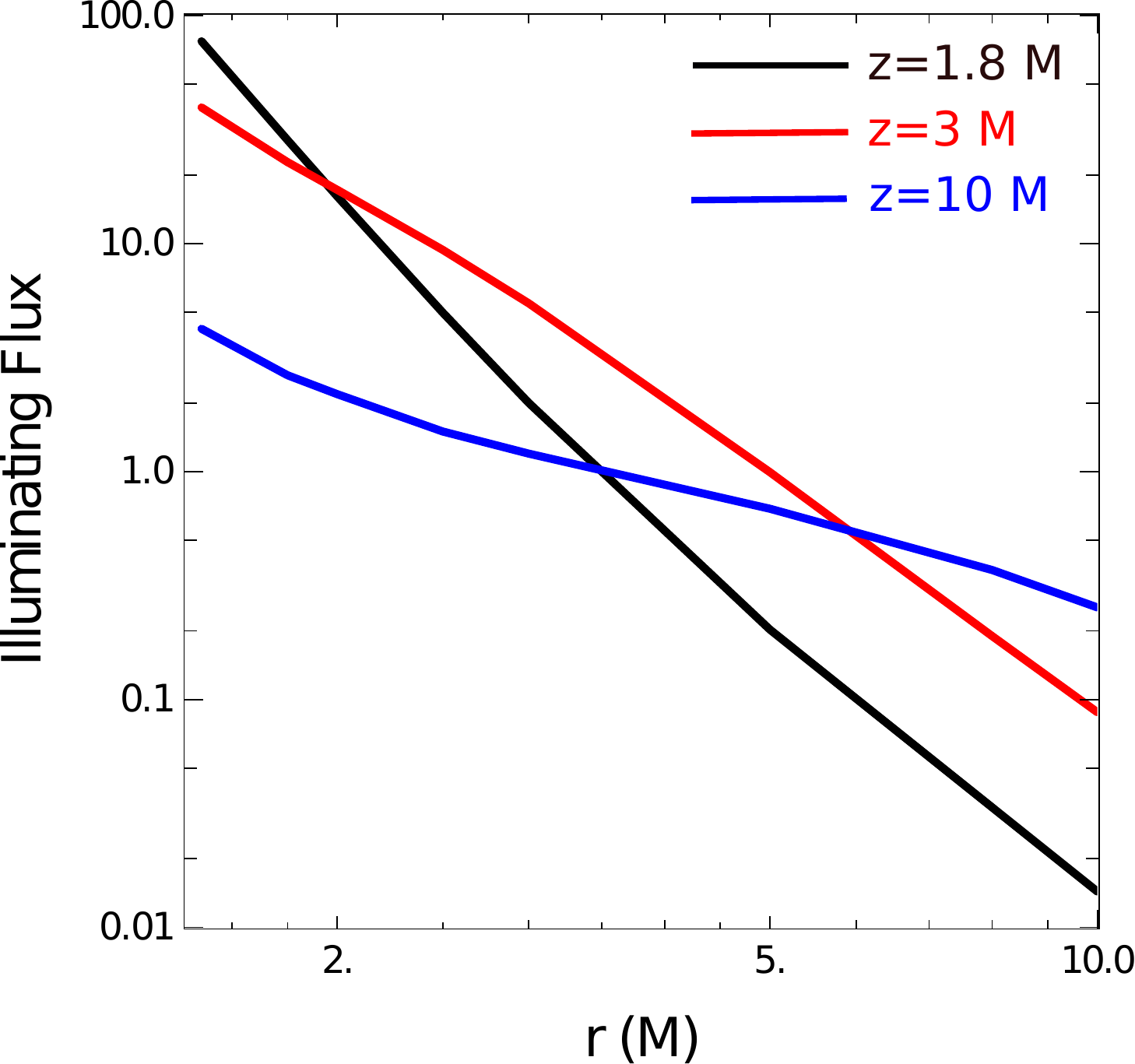}
\includegraphics[width=6cm,height=6cm]{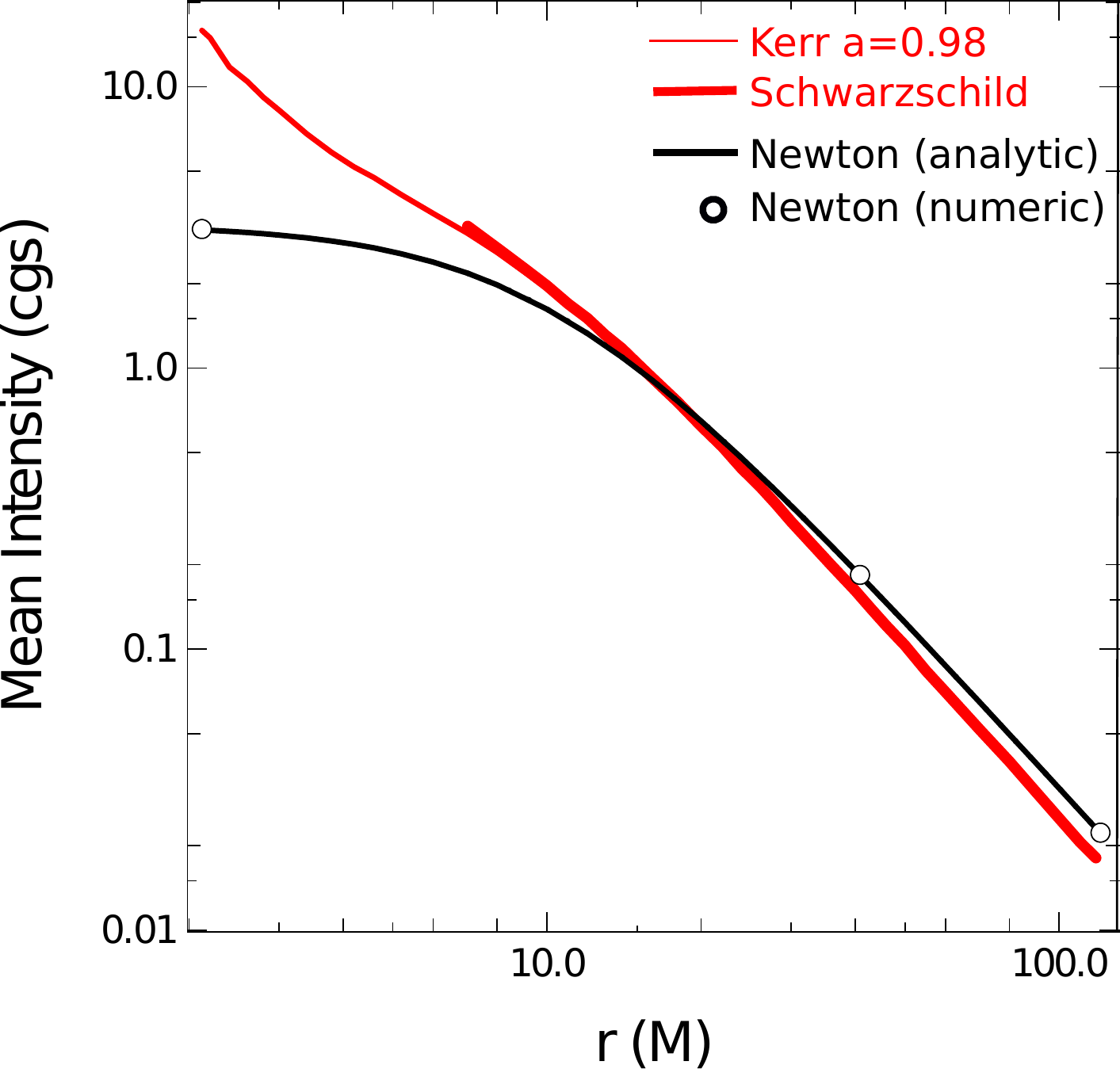}
\caption{\textbf{Left:} Illuminating flux, $\mathcal{F}(r)$, as defined in Eq.~\ref{eq:illflux}
for a spectral index $\alpha=1$ and three different altitudes of the lamp,
$z=1.8\,M$ (magenta), $z=3\,M$ (red) and $z=10\,M$ (blue). Radius is in
$M$ units and flux in arbitrary units. This figure
is extremely similar to Fig.~2, lower panel, of~\citet{dauser13}.
\textbf{Right:} Illuminating mean intensity in cgs units, $\mathcal{I}(r)$, 
computed in a Newtonian spacetime (solid black
and circles),
a Schwarzschild spacetime (thick red) and a Kerr spacetime with spin parameter $a=0.98$ (thin red).
The black circles are Newtonian illuminating fluxes as computed by GYOTO while the solid black line
is the analytical expression of the Newtonian mean intensity.}
\label{fig:compare}
\end{figure*}

Let us now derive the illuminating flux in the direction normal to the disk,
in order to properly define the normalizing constant $A$.
%{which is important for comparing} our results to previous works. 
It reads
\bea
\label{eq:illflux}
F^{\mathrm{disk}}_\nu(r) &=& \int I^{\mathrm{disk}}_\nu \cos i \,\dd \Omega \\ \nn
&=& {A} \nu_{\mathrm{disk}}^{-\alpha} \mathcal{F}(r) \\ \nn
\eea
where we introduce the quantity $\mathcal{F}(r) = \int g^{3+\alpha}\cos i \,\dd \Omega$
which is independent of the frequency. Figure~\ref{fig:compare}, left panel, shows the evolution
of this quantity with radius for three different small values of the lamp altitude $z$ and for
a spectral index of $\alpha=1$. This Figure shows a perfect agreement with Figure~2,
lower panel, of~\citet{dauser13}. We consider this test as a consistency check with
previous works.
We are able to compute precisely illuminating fluxes or mean intensities
for any kind of lamp position and geometry. We are also able to consider either 
a black-hole metric, or flat spacetime in order to compare Newtonian and general-relativistic (GR)
illuminating fluxes or mean intensities. Newtonian illuminating mean intensities are computed by 
assuming a Minkowski metric, an observer at rest at coordinate radius $r$, and fixing to $g=1$ the redshift
factor (in order to remove all relativistic effects including special relativistic effects that would
still be present in a Minkowski spacetime if $g$ is not fixed to $1$).

At this stage, we only need to define the normalizing constant $A$ in order
to compute mean intensities $\mathcal{I}(r)$ in cgs units. We do this by choosing 
 one particular value of the total
X-ray luminosity illuminating an accretion disk. This total luminosity can be
written
\be
L_{X,\mathrm{illum}} = 2\int_{\nu_{\mathrm{lamp}}\in X}\int_{\mathcal{S}_{\mathrm{disk}}} F^{\mathrm{disk}}_\nu(r) \dd S_{\mathrm{disk}}(r) \dd \nu_{\mathrm{disk}}(r)
\ee
where $\nu_\mathrm{lamp}$ is the lamp-frame frequency which is varied in the X band between
$\nu_{\mathrm{lamp,1}}$ and $\nu_{\mathrm{lamp,2}}$, $\nu_{\mathrm{disk}}(r)$ is the corresponding
redshifted frequency at radius $r$ in the disk frame, and $\dd S_{\mathrm{disk}}(r)$ is the infinitesimal
disk area between $r$ and $r+\dd r$ in the disk frame. 
%{The factor of $2$ in this expression comes from the fact that if the lamp is the base of a jet,
%it should be present on both sides of the disk, thus both faces of the disk should be considered.}
\myred{The factor of $2$ in this expression is justified by considering a symmetric source,
illuminating both sides of the disk.
We note that our numerical treatment only considers illumination from one
side of the disk. However, in a realistic situation (for instance if the source is the
base of a jet), this illuminating source should be present
on both sides of he disk. The luminosity $L_{X,\mathrm{illum}}$ above is defined in order to normalize all quantities
to realistic values. As a consequence, it should contain this factor of $2$. Considering two symmetric sources
in the numerical treatment would not change anything because the observer will always see only one side
of the disk (neglecting the small contribution of the very gravitationally bent radiation coming from the lower side that can be seen from
the upper side of the disk).}
In the Kerr metric, the redshift factor $g$ is known
analytically so that we may write in the equatorial plane
\be
\nu_{\mathrm{disk}}(r) = \nu_{\mathrm{lamp}} \frac{r^{3/2}+a}{\sqrt{r^3+2ar^{3/2}-3r^2}} \sqrt{\frac{z^2+a^2-2z}{z^2+a^2}}
\ee
where $a$ is the black hole spin parameter.
The element of area in the equatorial plane is 
\bea
\dd S_{\mathrm{disk}}(r) &=& \sqrt{g_{rr} g_{\pp\pp}} \dd r \dd \pp \\ \nn
&=& 2 \pi r \sqrt{\frac{r^2 + a^2 + 2 a^2 / r}{r^2 - 2 r + a^2}} \dd r\\ \nn
\eea
where $g_{rr}$ and $g_{\pp\pp}$ are Kerr metric coefficients.
The total illuminating luminosity is thus
\bea
L_{X,\mathrm{illum}}  &=& 2A \frac{ \nu_{\mathrm{lamp,2}}^{1-\alpha} - \nu_{\mathrm{lamp,1}}^{1-\alpha}}{1-\alpha}  \\ \nn
&&\times  \int_{r_{\mathrm{min}}}^{r_{\mathrm{max}}} \mathcal{F}(r) 2 \pi r \sqrt{\frac{r^2 + a^2 + 2 a^2 / r}{r^2 - 2 r + a^2}} \\ \nn
&&\times \left(\frac{r^{3/2}+a}{\sqrt{r^3+2ar^{3/2}-3r^2}} \sqrt{\frac{z^2+a^2-2z}{z^2+a^2}} \right)^{1-\alpha} \dd r \\ \nn
\eea
which can be integrated numerically. Here, $r_{\mathrm{min}}$ and $r_{\mathrm{max}}$ 
are the inner and outer radii of the accretion disk. Thus, by choosing $L_{X,\mathrm{illum}} $, the normalizing
constant $A$ is known. We note that this normalization depends on the spin parameter.
In order to consider one common lamp for all simulations presented in this work, we
choose to normalize the problem for $a=0$ and to use the same normalization whatever the spin.
This means that $L_{X,\mathrm{illum}} $ will not be the same for different spins, it is rather the lamp which is
kept the same for all simulations.

%Let us introduce the X-band luminosity
%\bea
%L_{X,\mathrm{illum}}  &=& \int_{\nu_{\mathrm{disk}}\in X}\int_{\mathcal{S}(r_o)} F^{\mathrm{disk}}_\nu(r_o) \,\dd S \,\dd \nu_{\mathrm{disk}} \\ \nn
%&=& A \frac{ \nu_{\mathrm{disk,2}}^{1-\alpha} - \nu_{\mathrm{disk,1}}^{1-\alpha}}{1-\alpha}  \,4 \pi r_o^2 \,\mathcal{F}_{\mathrm{N}}(r_o)\\ \nn
%\eea
%where $\mathcal{S}(r)$ is the surface of a sphere of coordinate radius $r$,
%$\nu_{\mathrm{disk,1}}$ and $\nu_{\mathrm{disk,2}}$ are the limiting X-band frequencies, and
%$r_o$ is the outer disk radius (expressed in cgs units in the expression above).
%The quantity $\mathcal{F}_{\mathrm{N}}(r_o)$ is the Newtonian illuminating flux
%at the disk outer radius, which will be used to normalize all simulations (including simulations
%with GR illumination) in order to be able to compare them.
%From the above equation, 
%the constant $A$ can be computed in cgs units once the X-band luminosity is specified.
For a spin-0 black hole, we consider a luminosity of {$L_{X,\mathrm{illum}} (a=0)=10^{36}\,\mathrm{erg}\,\mathrm{s}^{-1}$.
This choice fixes the normalizing constant $A$, which is kept the same for both $a=0$
and $a=0.98$. The corresponding illuminating luminosity for $a=0.98$ is $L_{X,\mathrm{illum}} (a=0.98) = 1.3\times10^{36}\,\mathrm{erg}\,\mathrm{s}^{-1}$.
We note that the accretion-related luminosity is equal $L_{X,\mathrm{acc}} = 2\times10^{37}\,\mathrm{erg}\,\mathrm{s}^{-1}$
with our choice of accretion rate and taking an accretion efficiency of $\eta=0.1$. This means that the accretion luminosity
is bigger than the illumination by an order of magnitude.
\red{Consequently, we are modeling a thermally dominated state with a weak hard tail~\citep[for a review on X-ray binaries states
and radiative processes, see][]{zdziarski04}.}
}

We keep constant the following parameters:
$z=10\,M$, $\alpha=0.7$,
$h \nu_{\mathrm{lamp,1}} = 2$~eV and $h \nu_{\mathrm{lamp,2}}=10^5$~eV.
All these fixed parameters are also given in Table~\ref{tab:param}.
In particular, we will not vary the lamp height $z$ although this parameter has an important
impact on the illumination~\citep[see e.g.][]{dauser13}. Our goal here is not to scan the
full parameter space but rather to determine the effect of the spin parameter on the reflected
spectrum in a setup where the lamp is sufficiently close to the black hole to lead to
{non-negligible} light-bending effects.
%{\bf to lead to clear strong-field effects} {\it I do not understand this sentence}.
\begin{table}[htbp!]
\centering \caption{Lamp post model fixed parameters used in this article.}
\begin{tabular}{l{c}{c}}
Parameter                  &      Notation       & Value                                \\
\hline
Black hole mass                       & $M$         &          $10\,M_{\odot}$                             \\
Accretion rate                       & $\dot{m}$         &          $0.01\,\dot{m}_{\mathrm{Edd}}$                             \\
Lamp height                & $z$         &        $10\,M$        \\
Spectral index                & $\alpha$         &        $0.7$        \\
Disk inner radius           & $r_{\mathrm{min}}$   &              $7M (a=0) / 2M (a=0.98)$                     \\
Disk outer radius           & $r_{\mathrm{max}}$   &             $120\,M$                     \\
Illumination bounds & $h \nu_{\mathrm{lamp}}$& [2 eV ; $10^5$ eV] \\
%Luminosity             & $L_{X,\mathrm{illum}} $ & $5\times10^{35}\,\mathrm{erg}\,\mathrm{s}^{-1}$                 \\
\end{tabular}
\label{tab:param}
\end{table}

The resulting mean intensities in cgs units,
considering Newtonian, Schwarzschild and close-to-extreme Kerr spacetimes
are presented in \red{Fig.~\ref{fig:compare}, right panel}.
This Figure in particular shows a comparison between the GYOTO-computed 
Newtonian mean intensity and its analytical expression 
\be
\mathcal{I}_{\mathrm{Newton}} \propto \frac{1}{r^2 + z^2}.
\ee
The agreement is within $\approx 0.2 \%$. We note that even at large distances, the illuminating 
relativistic mean intensities differ (by roughly $25\,\%$) from the Newtonian ones. This
can be understood by noticing that even when the Keplerian observer is at a large
distance, the illuminating radiation is always emitted in the strong gravitational 
field region and thus always contains a relativistic signature.
%\begin{figure*}[htbp]
%\centering
%\includegraphics[width=6cm,height=6cm]{MeanIntensitiesCgs.pdf}
%\caption{Illuminating mean intensity in cgs units, $\mathcal{I}(r)$, 
%computed in a Newtonian spacetime (solid black
%and circles),
%a Schwarzschild spacetime (thick red) and a Kerr spacetime with spin parameter $a=0.98$ (thin red).
%The black circles are Newtonian illuminating fluxes as computed by GYOTO while the solid black line
%is the analytical expression of the Newtonian mean intensity.}
%\label{fig:meanintens}
%\end{figure*}

At this stage, we are ready to compute the reflected spectrum due to the
reprocessing of the lamp radiation by the disk. 
We will consider an illuminating mean intensity
as computed in a Newtonian spacetime, or in a Kerr spacetime with spin parameter $a=0$ or $a=0.98$.
%{\it why only one spin is listed here? this is missleading, maybe we should remove 
%the formula here, or add $a=0$ example}.

\section{Reflected spectra}

\subsection{Local spectra}
\label{sec:localspec}

Local spectra are computed using the radiative transfer code ATM21 as 
described in~\citet{rozanska11}. We refer to this paper, as well as to~\citet{madej04} and~\citet{rozanska08}, 
for more details and will give here only the most important steps.
\blue{We stress that ATM21 computes
simultaneously the structure of an irradiated atmosphere and its outgoing line
and continuum spectrum, the atmosphere being both in hydrostatic and radiative 
equilibrium}.

Our radiative transfer equation assumes a plane-parallel geometry and reads
\be
\mu \frac{\dd I_\nu}{\dd \tau_\nu} = I_\nu - \frac{j_\nu}{\kappa_\nu + \sigma_\nu}
\ee
where $\mu = \cos i$ is the cosine of the angle between the light ray and the local normal, 
$j_\nu$ is the emissivity, $\kappa_\nu$
is the opacity, $\sigma_\nu$ is the scattering coefficient,
and $\dd\tau_\nu = -(\kappa_\nu+\sigma_\nu) \rho \,\dd z$, where $\rho$ is the mass density, is the optical depth.
Emissivity takes into account the disk thermal emission, Compton scattering
redistribution functions and iron fluorescence lines from the gas at 
each ionization state. 

In our numerical procedure, the equation above is solved with the structure of the gas kept  
in \blue{radiative}, hydrostatic and ionization equilibrium. We use local thermal equilibrium (LTE) 
absorption $\kappa_\nu$, whereas coefficients of emission $j_\nu$ and scattering $\sigma_\nu$
include non-LTE terms. 
The coefficient of true absorption considers photoionization from numerous levels
of atoms and ions as well as Bremsstrahlung (free-free) absorption from all ions. 

The external illumination by X-ray photons is fully taken into account as an additional 
intensity field which enters the atmosphere from above. This energy-dependent radiation 
modifies the temperature and the gas ionization level in addition to the disk thermal 
radiation generated below the atmosphere via viscosity mechanism. 

Both radiation intensities are scattered by Compton process, 
which is included in the radiative transfer equation using angle-averaged Compton redistribution functions 
\citep[after][]{pomraning73,guilbert81,kershaw87}. Compton scattering cross-sections
were computed following the paper by~\citet{guilbert81}, \red{corrected for computational errors
in the original paper}~\citep[see][]{madej16}, and for the first time presented in a 
computer code by~\blue{\citet{madej89}}.
Our equations and the Compton redistribution functions work correctly in cases of both 
large and small energy exchange between X-ray photons and free electrons at the time of scattering.
They ensure accurate solution of the radiative transfer also in cases when the initial photon energy
before or after scattering exceeds the electron rest mass ($m_{\rm e}c^2=511$~keV).

We highlight that we compute the full reflected spectra, meaning that we take into
account both the thermal component emitted by the accretion disk at the local
temperature and the reflected part due to the reprocessing of the light emitted by the lamp.
Therefore, the disk emission and X-ray illumination always self-consistently influence the matter 
thermal and ionization structure. As a result, the outgoing local spectrum contains all reprocessing 
signatures, such as: 
the deviation of the thermal spectrum from pure blackbody due to photo-absorption and re-emission, 
the deviation of the spectrum due to photon energy shift by Compton scattering, 
the Compton hump due to the reflection from the heated atmosphere
\citep[both effects are presented in][]{madej2000}, 
and the fluorescent iron line complex \citep{rozanska08,rozanska11}.

\subsubsection{Computing local spectra}

We model the accretion disk by
a stationary relativistic slim disk solution \citep{sadowski11}, which depends on the black hole mass and assumed accretion rate, assumed here to be: $M=10$~M$_{\odot}$ and $\dot{m} = 0.01\,\dot{m}_{\mathrm{Edd}}$.  
This allows to compute the vertical gravity and effective temperature radial distributions that are
input parameters of our radiative transfer computations. These quantities depend on the spin parameter, which is
thus taken into account self-consistently. From those quantities, 
the radiative transfer is solved for a set of discretized radii, 
assuming that the disk has solar-like chemical abundances. At each radius, vertical gravity and 
effective temperature are only needed to formulate the boundary conditions of the vertical atmosphere's structure.  
Subsequent iterations proceed between the radiation field and the gas structure, as it should be when solving radiation transfer problem. 
For the atmosphere, we assume hydrostatic \blue{and radiative} equilibrium, which adds an additional differential equation to the whole problem, making
the derivation of the solution 
much more complex and time consuming than in the case of a constant-density gas~\citep[as done in][]{garcia14}.

The Fe $K\alpha$ doublet fluorescence 
lines are set to central energies from $6.404$~keV and $6.391$~keV to 
$6.652$~keV and $6.639$~keV, depending on the matter ionization level. 
The iron $K\beta$ line energy centroid is $7.057$~keV. Their natural
widths are set respectively to $2.7$~eV, $3.3$~eV and $2.5$~eV.
The boundary condition of the radiative transfer equation is set by the illumination
mean intensity profile computed in Section~\ref{sec:illum} which gives the value
of the mean intensity for all radii and the associated direction cosine $\mu_{\mathrm{illum}}$.
Deep in the disk atmosphere, we assume full thermalization. Nevertheless, 
in our scheme when iterating between the gas structure and the radiation field at each point of the 
atmosphere, including the deepest one, temperature corrections are taken into account.

The radiative transfer equation is solved for a set of $\approx 30$ radial points,
$8$ directional cosines $\mu$
between $0.02$ and $0.98$ and for $\approx 2600$ values of photon energy
between $2$~eV and $10^5$~eV. 
The precise definition of the grids considered
for the two spin values we take into account is given in Table~\ref{tab:grid}.
The output of this Section is the value of the reprocessed specific intensity
$I_\nu^{\mathrm{repro}}$ as a function of the $8$ direction cosines $\mu$
and of the $\approx 2600$ photon energies, for all values of radii in the disk where 
the radiative transfer is solved. 
This final reprocessed intensity consists in both the thermal emission from the cold accretion disk and 
the mixture of absorbed and then re-emited radiation, either by true absorption or 
scattering, outgoing from the hot skin. 
\begin{table}
\centering \caption{Local spectra are computed at $n_r$ radial points chosen between
$r_{\mathrm{min}}(a)$ and $r_{\mathrm{max}}$, where $a$ is the spin parameter. The radial grid is denser close to
$r_{\mathrm{min}}$ and becomes more sparse as $r$ increases. 
When two values are given separated by a $"/"$ sign, the first one is used
for $a=0$ and the second one for $a=0.98$.
The ISCO radius for both spins are $r_{\mathrm{ISCO}}(a=0)=6\,M$ and
$r_{\mathrm{ISCO}}(a=0.98)=1.6\,M$. The smallest radial value is always
chosen slightly above the ISCO.
The grid contains $n_i$ direction cosines $\cos i$ and $n_E$
photon energies.}
\begin{tabular}{lc}
                  &                     \\
\hline
$r_{\mathrm{min}},r_{\mathrm{max}},n_r$                       & $7 M / 2 M, \,120 M, \,31 / 36$                         \\
$n_i$                & $8$                  \\
$\cos i$ &$(0.02,0.1,0.2,0.4,0.6,0.75,0.9,0.98)$ \\
$n_E$           & $2600$                     \\
\hline
\end{tabular}
\label{tab:grid}
\end{table}

%\begin{table}
%\centering \caption{Local spectra are computed at $n_r$ radial points chosen between
%$r_{\mathrm{min}}(a)$ and $r_{\mathrm{max}}$, where $a$ is the spin parameter. The radial grid is denser close to
%$r_{\mathrm{min}}$ and becomes more sparse as $r$ increases. This Table contains
%only the spin-independent, sparse radial grid values.
%The ISCO radius for both spins are $r_{\mathrm{ISCO}}(a=0)=6\,M$ and
%$r_{\mathrm{ISCO}}(a=0.98)=1.6\,M$. The smallest radial value is always
%chosen slightly above the ISCO.
%The grid contains $n_i$ direction cosines $\cos i$ and $n_E$
%photon energies.}
%\begin{tabular}{l{c}c}
%                  &   $a=0$           &             $a=0.98$                   \\
%\hline
%$r_{\mathrm{min}},r_{\mathrm{max}},n_r$                       & $7 M, \,120 M, \,9$         &          $2 M, \,120 M, \,17$                  \\
%Sparse radial grid &  \multicolumn{2}{c}{$(12M,16M,20M,40M,60M,80M,120M)$} \\
%$n_i$                & $8$         &       $8$         \\
%$\cos i$ & \multicolumn{2}{c}{$(0.02,0.1,0.2,0.4,0.6,0.75,0.9,0.98)$} \\
%$n_E$           & $2600$   &           $2600$                        \\
%\end{tabular}
%\label{tab:grid}
%\end{table}

\subsubsection{Dependence on illumination computation, spin and direction}

\begin{figure*}[htbp]
\centering
\includegraphics[width=15cm,height=10cm]{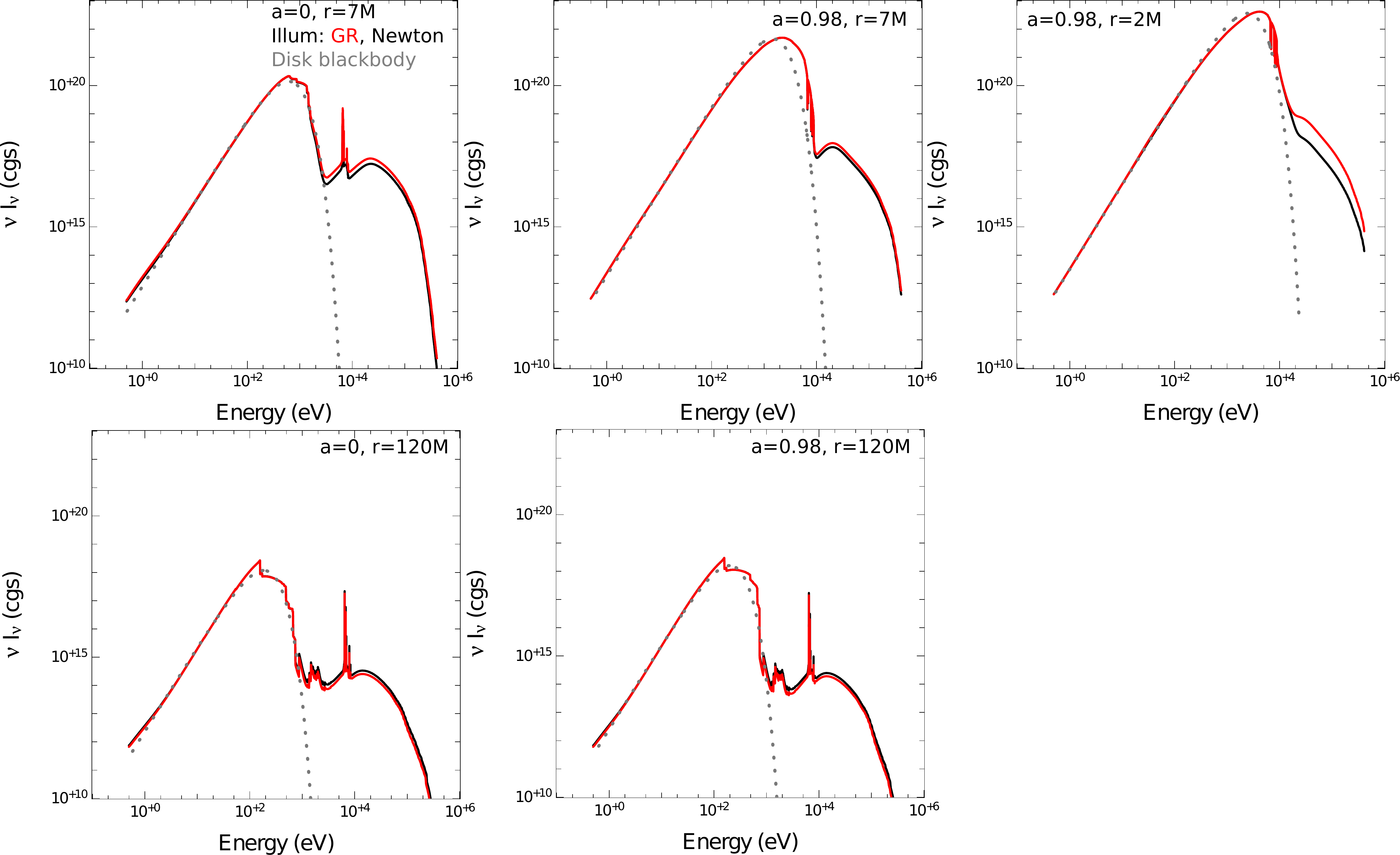}
\caption{{\it Local $\nu I_\nu$ spectra: impact of illumination computation (Newtonian or GR) and spin}.
Only one direction cosine is considered, $\cos i = 0.98$ (almost face-on, i.e. close to the normal to the disk).
The illumination is computed in general relativity (red) or in a Newtonian spacetime (black).
The grey dotted spectrum is the blackbody function at the local temperature.
The spin is $0$ is the left column and $0.98$ in the right column.
The spectrum is computed at $r=2M$ (upper row, right panel), $r=7M$ (upper row left and middle panels) or $r=120M$ (lower row).}
\label{fig:localspectraIllumSpin}
\end{figure*}

\begin{figure*}[htbp]
\centering
\includegraphics[width=10cm,height=5cm]{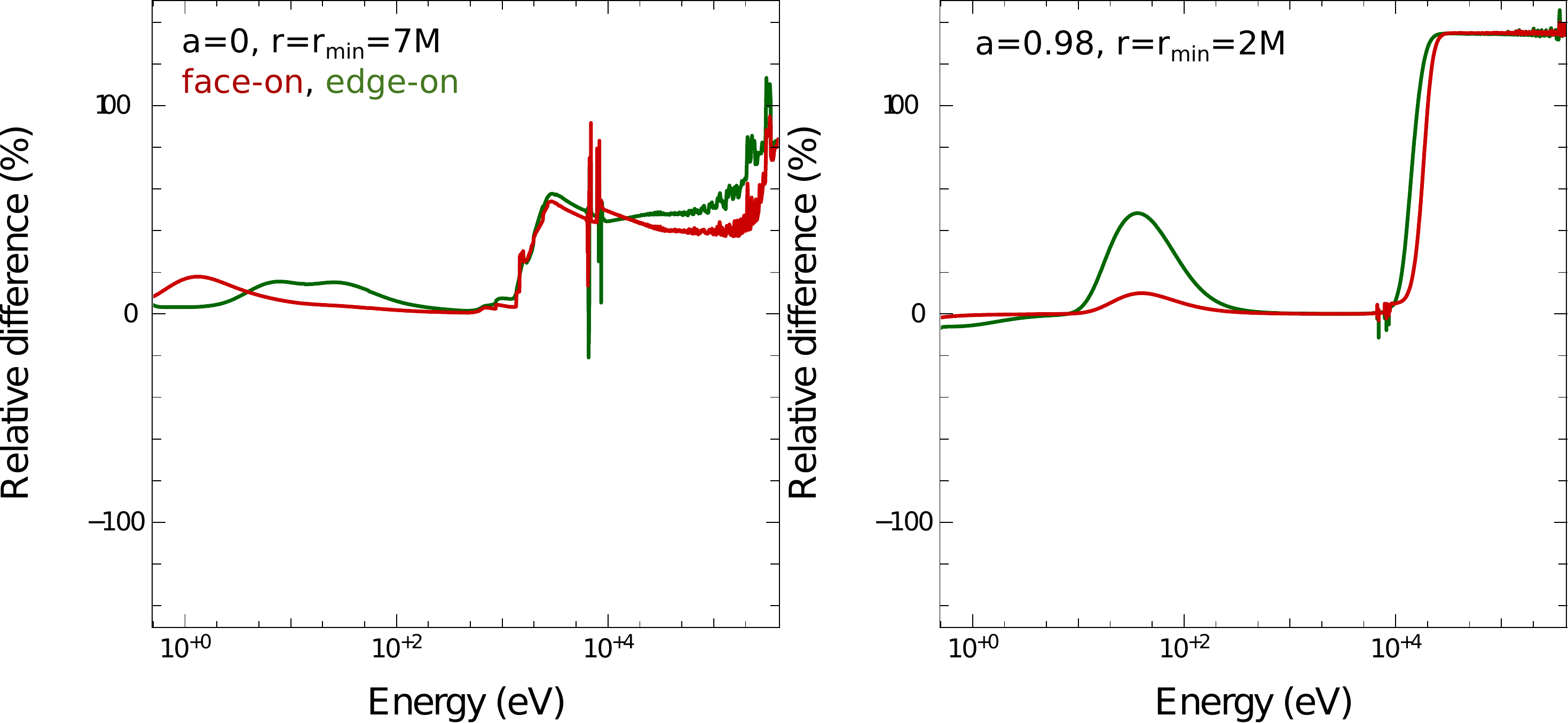}
\caption{{\it Local $\nu I_\nu$ spectra: relative difference at $r_{\mathrm{min}}$ for GR/Newtonian illuminations}.
The red curve shows the relative difference between the local spectrum computed with GR or
Newtonian illuminations in the face-on direction. The green curve shows the same quantity for
the edge-on direction. Spin is $0$ on the left panel and $0.98$ on the right panel. The spectrum
is computed in both cases at $r=r_{\mathrm{min}}$. %\textbf{Right: }local edge-on spectrum for spin $a=0.98$
%at $r=2M$ (same legend as Fig.~\ref{fig:localspectraIllumSpin}).
}
\label{fig:relativediffillum}
\end{figure*}

Fig.~\ref{fig:localspectraIllumSpin} illustrates the effect of changing the computation of
the illumination (Newtonian or GR) and the value of spin on
the local spectrum. As illustrated in the upper left panel, the spectrum is made of
three parts: 
\begin{itemize}
\item the thermal bump (around $100\,\mathrm{eV} - 1\,\mathrm{keV}$) is %the blackbody spectrum 
due to two forms of dissipation: viscous dissipation in the accretion disk and heating by
the irradiation of the lamp. {However for our choice of parameter, the viscous heating
is dominating (see the end of Section~\ref{sec:illum})}; %the blackbody spectrum at the hydrodynamical equilibrium
%temperature of the disk and the reflected blackbody due to the lamp irradiation
%\textbf{Agata, could you please check that this sentence is fully correct, and
%change it if needed?};
%This part of the disk is not due to the reflection by the
%disk of the lamp illumination. The reflected spectrum is made of the two following parts,
\item the iron-line complex due to fluorescent emission (around $6.4$~keV);
\item and the Compton hump (at energies just above the iron-line complex) due
to the Compton back scattering of hard photons.
\end{itemize}
The thermal part of the spectrum will be less affected by changing the illumination
as part of it is computed from the hydrostatic equilibrium temperature of the
accretion disk which only slightly depends on illumination.  
The reflected part of the spectrum (iron-line complex plus
Compton hump) depends on the illumination profile considered.
At small radii, the reflected spectrum
is stronger for GR illumination, while at bigger radii, the opposite is true and the Newtonian
illumination leads to a stronger reflected spectrum. This is in perfect agreement with
Fig.~\ref{fig:compare}, right panel, which shows that the Newtonian illumination dominates the
general-relativistic one at big radii, while it is smaller at small radii, because of
light bending effect.
Fig.~\ref{fig:relativediffillum} shows the relative difference $\mathcal{R}$ between the local spectra
computed with a Newtonian or GR illumination, at the inner disk radius (where most
of the radiation is emitted). It is computed following
\be
\mathcal{R} = 200 \frac{\mathrm{spectrum\, GR} - \mathrm{spectrum \,Newton}}{\mathrm{spectrum\, GR} + \mathrm{spectrum\, Newton}}.
\label{eq:reldiff}
\ee
%\textbf{For the time being there's an abs() at the numerator, I will remove it later.}
Fig.~\ref{fig:localspectraIllumSpin} also shows the influence of the spin parameter on the local
spectrum. Higher spin leads to higher disk temperature~\citep[see Fig.~1 of][]{rozanska11},
particularly at smaller radii (the spin dependence decreases to zero as $r$ increases).
As a consequence, at $r=7M$, the thermal bump
peaks higher and so does the Compton hump as Comptonization is depending on the disk
temperature. At $r=120M$ the spectra are very similar for $a=0$ and $a=0.98$,
as they should be. 
The thermal bump for spin $a=0.98$ deviates from blackbody radiation since for the inner regions 
the disk effective temperature is high. Such an atmosphere is made of an almost completely ionized gas allowing for 
efficient Compton scattering. This modifies the high energy tail \blue{and the frequency of peak flux} of thermal radiation as 
demonstrated in~\citet{madej91}.
The outgoing spectrum is sensitive to the local vertical temperature and ionization structure. 
Therefore, spectra emerging from the outer radii at r=120~M for both spins may sligthly differ, 
since reprocessing depends on many aspects %starting from the exact illumination field and on the detailed 
like the exact illumination field and the detailed 
vertical structure. % ending. 
The appearance of spectral features also depends on local thermodynamical conditions. 
If the external illumination is low in %addition to the disk temperature, 
comparison to the disk temperature, 
we clearly see ionization edges in absorption. But when the illumination increases, many spectral features 
start to be visible in emission as shown by~\citet{madej2000}.
When matter is completely ionized we do not see any spectral features in the reflected 
part of the spectrum. 
Finally, if the inner disk temperature is high enough, thermal radiation modified by Compton 
scattering covers the iron line region, and this feature is only visible as the resonant line in 
absorption from the hot atmosphere~\citep{rozanska11}. 
This result clearly shows that properly computed disk thermal component should be taken into account in the models used for spin fitting 
from iron line shape in black hole binaries. 
 
%At small radius, the main difference is in the thermal bump. It is higher by a factor
%of $\approx 50$ for $a=0.98$. This is due to the difference in effective temperature of the
%equilibrium disk depending on the spin value. Fig.~1 of~\citet{rozanska11} shows the evolution
%of the effective temperature at the two values of spin considered here. It is bigger by a factor
%of $\approx 7$ at $r=6M$ for spin $a=0.98$ as compared to $a=0$. \textbf{To be understood:
%this temperature diff does not seem to be of the same order as the Bnu diff. Also comment
%on the difference of Comptonization at r=6M. Also comment on iron line in em or in abs depending
%on spin, i.e. on the thermal bump.}. 

\begin{figure*}[htbp]
\centering
\includegraphics[width=15cm,height=10cm]{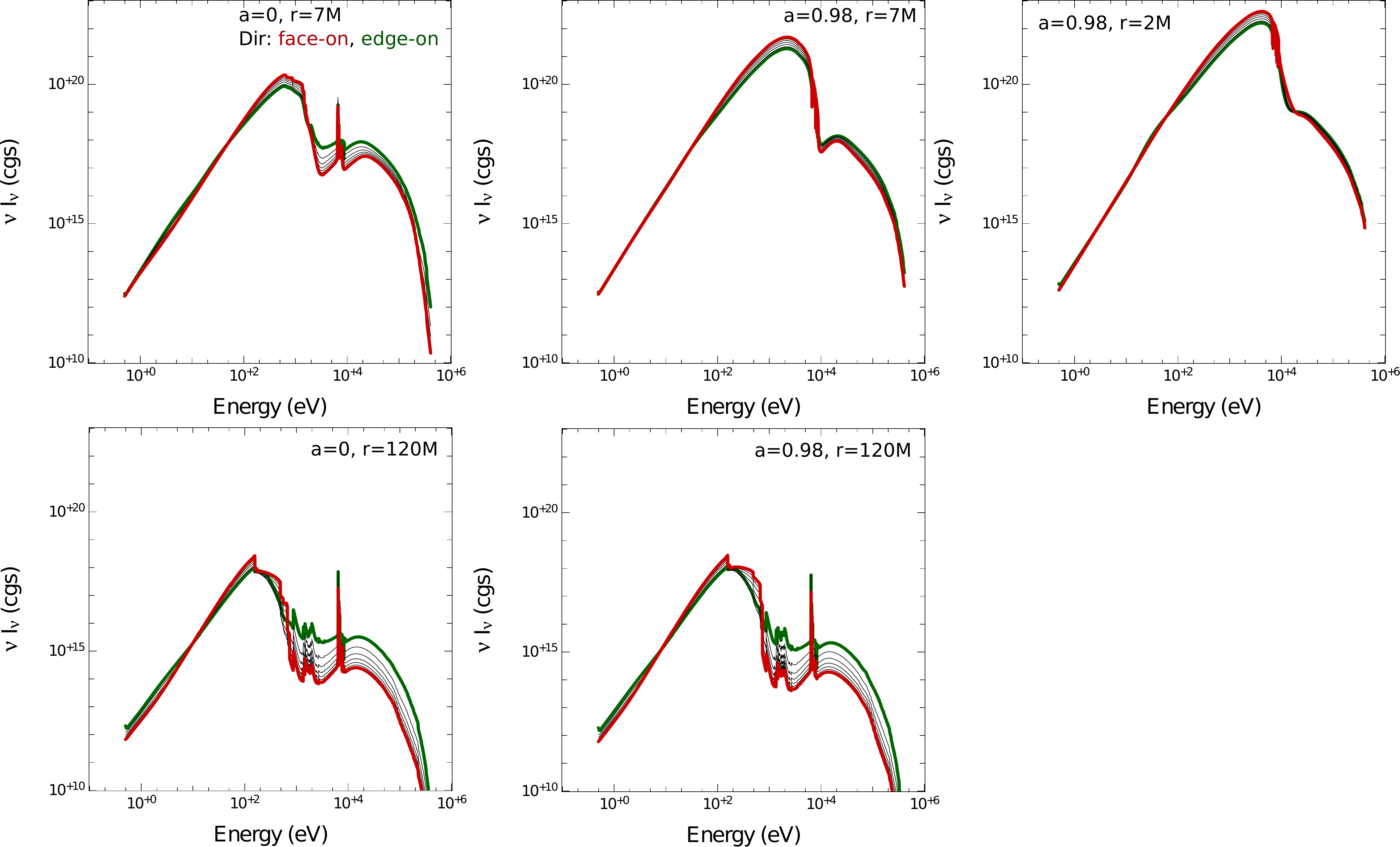}
\caption{{\it Local $\nu I_\nu$ spectra: impact of direction}.
The illumination is computed in general relativity.
The spectrum is computed at $r=7M$ (top row) and $r=120M$ (bottom row).
The spin is $0$ (left panel) or $0.98$ (right panel).
All $8$ direction cosines $\cos i$ given in Table~\ref{tab:grid} are represented with almost face-on reflected
spectrum ($\cos i = 0.98$) in red and edge-on ($\cos i = 0.02$) in green.
}
\label{fig:localspectraIllumDirec}
\end{figure*}

Fig.~\ref{fig:localspectraIllumDirec} illustrates the dependence on the direction 
of the local spectrum. For both spins, edge-on ($\cos i = 0.02$) emission is smaller
in the thermal bump but higher in the reflected part of the spectrum. This is of course
in agreement with the discussion in~\citet{rozanska11} as we are using the same model (\red{but using
a denser radial grid}),
with only a differing illumination. It is interesting to note two more things.
First, for both spins, the reflected
part of the spectrum is more and more depending on direction as the radius gets bigger.
Second, at small radii, the reflected spectrum is much less dependent on direction
for higher spin than for lower spin. This is because the gas on innermost rings is hot itself 
and the thermal spectrum covers flat part of reflected spectrum. We see only steep part of the 
refleced spectrum which does not depend much on the direction.

\subsection{Ray tracing \red{observed} spectra}
\label{sec:raytrace}

At this point we are ready to start the third and last step of our numerical
pipeline, which is the ray tracing of the reflected spectrum to a distant observer.
%The distant observer is located at $r = 10^{10}\,M$ and photons are traced
%backwards in time from this observer towards the disk. 

{The main aim of this section is to analyze the importance of taking
into account the directional dependence (as a function of $\mu$) of the local
spectrum in order to compute observed spectra. 
\red{We note that we will
often use in this section the expression \textit{observed spectrum} having in
mind the reprocessed (thermal+reflected) spectrum ray traced to infinity. 
The direct component is shown separately to briefly discuss the relative strength of the
reprocessed/direct components, but we mainly focus in the following on the
reprocessed part.
}
This is the same analysis
as already done by~\citet{garcia14}, but with the important distinction of using
a very different code for computing the local spectrum.}

%\subsubsection{Emission angle dependence of ray-traced spectra}

Photons are ray traced backwards in time from an observer located at $r = 10^{10}\,M$.
When a photon hits the accretion disk at radius $r$, the disk-frame frequency $\nu$ 
and direction cosine $\mu$ are determined.
This last quantity is readily computed knowing the 4-velocity of the emitting
gas $\mathbf{u}$, the 4-vector tangent to the emitted
photon geodesic $\mathbf{p}$ and the local disk normal 4-vector $\mathbf{n}$
\be
\mu = -  \frac{\mathbf{n} \cdot \mathbf{p}}{\mathbf{u} \cdot \mathbf{p}}.
\ee
\begin{figure*}[htbp]
\centering
\includegraphics[width=16cm,height=5cm]{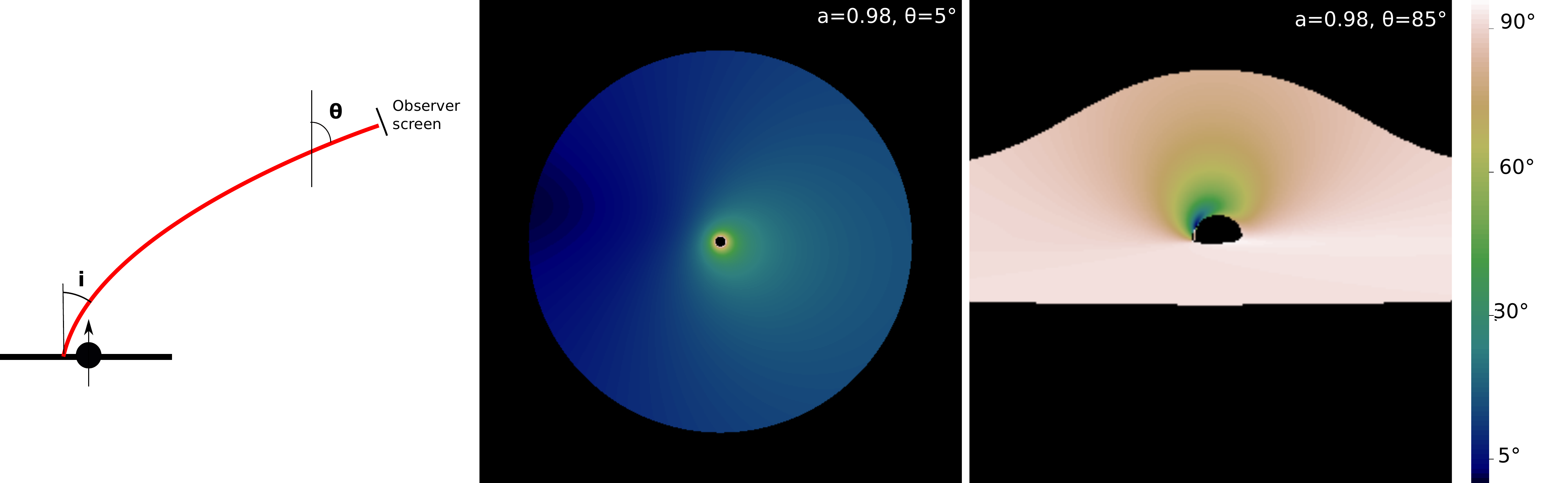}
\caption{{\it Map of emission angle for spin $a=0.98$}.
\textbf{Left:} geometry of the problem with the \textit{inclination angle} $\theta$
between the black-hole spin axis and the line of sight, and the \textit{emission angle}
$i$ between the disk normal and the direction of emission of the photon in the
emitter's frame.
\textbf{Center and Right:} 
these maps show a map of the values of the emission angle $i$ between the local normal
to the accretion disk and the direction of photon emission for a spin of $a=0.98$.
The inclination angle
is equal $\theta=5^\circ$ (center) or $\theta=85^\circ$ (right).
The right panel can be compared to Fig.~5, lower left panel of~\citet{garcia14}.
}
\label{fig:mapangle}
\end{figure*}
Fig.~\ref{fig:mapangle} shows the distribution of the emission angle $i$
(such as $\mu= \cos i$) in a disk surrounding an extreme Kerr black hole
seen under two different inclinations, $\theta=5^\circ$ and $\theta=85^\circ$.
Strong bending of light rays in the central region of spacetime leads to
a large set of emission angles being present for one given value of
the inclination $\theta$, in contrast to a Newtonian spacetime 
for which $i=\theta$. From this Figure only it is clear that considering
an emission-angle-averaged or a directional emission may lead to
very different observed spectra. It is the aim of this Section to investigate
this question, following the previous analysis by~\citet{garcia14}.
\begin{figure*}[htbp]
\centering
\includegraphics[width=8cm,height=7cm]{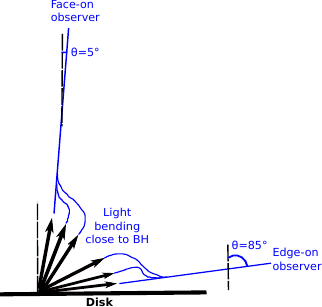}
\caption{{\it Illustrating the angle dependence of ray-traced spectra.}
Ray-traced spectra in this article are computed from two inclination angles,
$\theta=5^\circ$ (face-on observer) or $\theta=85^\circ$ (edge-on observer).
{From each of this inclination angles, photons are ray traced backwards to
the accretion disk (blue lines). For one given inclination $\theta$, photons reach the
disk with a set of different directions because of light bending. This is illustrated
in the Figure by the fact that one given blue line is connected to a set of different directions
of local emission in the disk (black arrows, the black vertical dashed line being the
local normal). This means that the face-on observer, e.g., sees photons
emitted not only at face-on emission but also in all other emission directions}. {\it Directional ray-traced spectra}
are computed by transporting along each geodesic the specific intensity $I_\nu(r,i)$ emitted
at the local position $r$ in the direction $i$ of this particular geodesic.
{\it Angle-averaged ray-traced spectra} are
computed by transporting along each geodesic the angle-averaged specific
intensity $I_\nu(r)$ obtained by averaging $I_\nu(r,i)$ over all emission angles $i$.
}
\label{fig:angleavg}
\end{figure*}

Fig.~\ref{fig:angleavg} illustrates the various angles used in our analysis
and defines in more detail the notion of angle-averaged ray-traced spectrum.
When a backwards-in-time ray-traced photon hits the disk, and after
having computed the emitted frequency $\nu$ and the direction cosine $\mu$,
the specific intensity $I_\nu^{\mathrm{repro}}(r,\mu,\nu)$ is trilinearly
interpolated {(i.e. linearly in all 3 dimensions)} from the results of Section~\ref{sec:localspec}. 
For computing an angle-averaged spectrum, this quantity is simply averaged
over the direction cosine $\mu$ 
\be
I_{\nu}^{\mathrm{repro,avg}}(r,\nu) = \int I_{\nu}^{\mathrm{repro}}(r,\mu,\nu) \dd \mu.
\ee
%\textbf{I have a small doubt here: I want to integrate over $\mu$, but why
%not over $i$? This wouldn't give the same result right?}

The specific
intensity in the distant observer's frame $I_\nu^{\mathrm{obs}}$ is then 
deduced by using the frame invariance of $I_\nu / \nu^3$. Thus a map
of specific intensity in the observer's frame (i.e. an image of the disk) can
be computed, which is readily transformed to a flux value by summing over
all directions on sky. The observed spectrum is thus at hand.

\begin{figure*}[htbp]
\centering
\includegraphics[width=11cm,height=15cm]{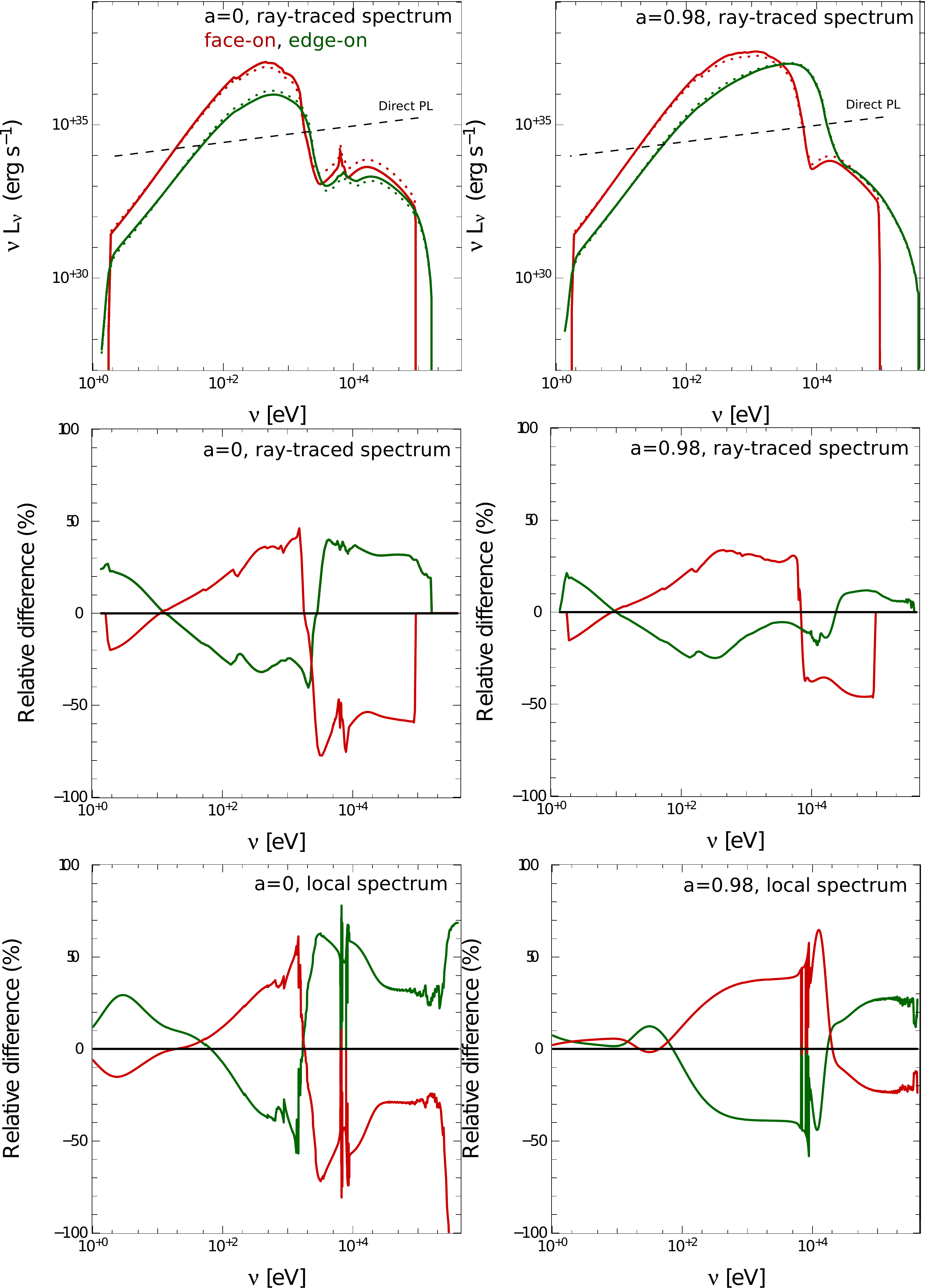}
\caption{{\it \red{Ray-traced} $\nu L_\nu$ spectra: impact of directionality}.
The local spectra have been computed with GR illumination.
The \textbf{top row} shows emission-angle-averaged (dotted) and angle-dependent (solid)
ray-traced spectra. \red{The straight black dashed line shows the level of
the direct-component power law (see text for details).} The \textbf{middle row} shows the relative difference
between averaged and angle-dependent ray-traced spectra.
The \textbf{bottom row} shows the same quantity as the middle row for the local spectra,
evaluated at $r=r_{\mathrm{min}}$.
The spin is $0$ (left column) or $0.98$ (right column).
Red curves refer to spectra ray traced from an inclination of $\theta=5^\circ$,
i.e. close to face-on.  Green curves are computed at an inclination of
$\theta=85^\circ$, i.e. close to edge-on. Note that "face-on" and "edge-on" in the
two upper rows (ray-traced quantities) refer to the inclination angle $\theta$, while in the bottom row (local quantities) 
these words refer to the emission angle $i$.
}
\label{fig:raytracedspectra}
\end{figure*}

Fig.~\ref{fig:raytracedspectra} shows the angle-averaged and
directional ray-traced spectra for both spins, together with their
relative difference. \red{This Figure also shows the level of the direct-component
power law, directly reaching the observer from the lamp.
%, which is not included
%in the ray-traced spectra. 
This component was computed by ray tracing the lamp
alone, as observed by the distant observer, and keeping the same normalization
of the emitted intensity as described in Section~\ref{sec:illum}. It shows that our
choice of parameters leads to a total spectrum dominated by the direct 
and thermal components.}
Fig.~\ref{fig:raytracedspectra} also shows the relative difference
between the directional and angle-averaged local (non-ray-traced) 
spectra, computed for both spins at $r=r_{\mathrm{min}}$. 
This Figure shows a few important things. As we will discuss this
quantity a lot, let us call $\mathcal{R}$ the relative
difference between angle-averaged and directional spectra,
which is computed in the same way as in Eq.~\ref{eq:reldiff}.
First, $\mathcal{R}$
can go as high as $\approx 70\%$, and in particular in the
region of the iron-line complex it is of the order of $50 \%$. This maximum value of $\mathcal{R}$ is the
same for local and ray-traced spectra. The general behavior of $\mathcal{R}$
is rather similar for local and ray-traced spectra. However, there is an important
difference: the edge-on value of $\mathcal{R}$ is always significantly smaller 
than its face-on value for ray-traced spectra (particularly at high spin), 
while for local spectra the difference is less pronounced. This means that
edge-on ray-traced spectra are closer to the angle-averaged spectra than
face-on ray-traced spectra. This fact can be understood as follows.
\begin{figure*}[htbp]
\centering
\includegraphics[width=10cm,height=5cm]{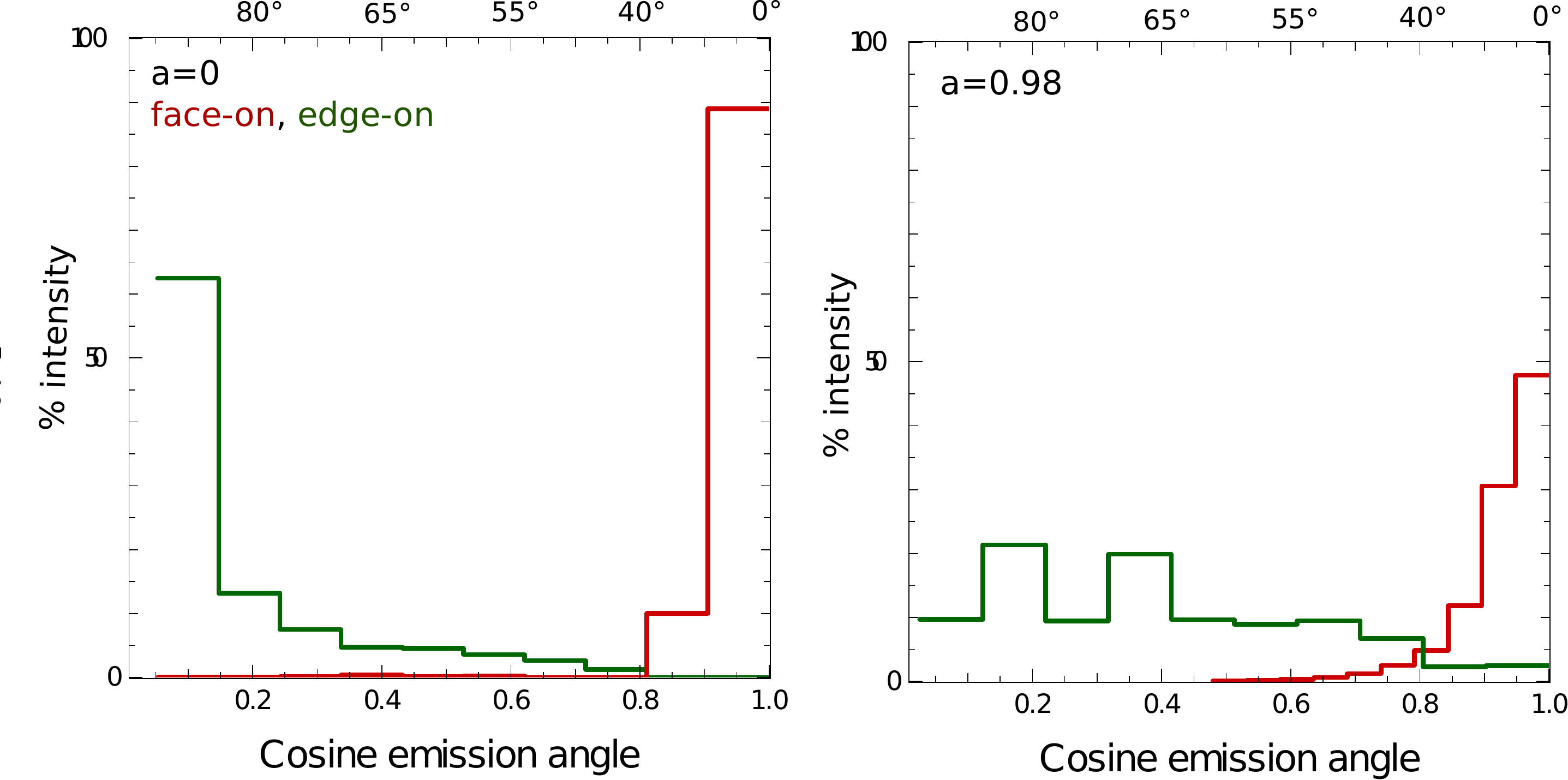}
\caption{{\it Specific intensity distribution with emission angle}.
This Figure shows the histogram of specific intensity emitted at the
disk's surface in percent of the total intensity as a function of $\cos i$. 
{This total intensity is defined simply as the sum of the specific intensity map
over all pixels. It is thus proportional to the total observed flux.
}
The spin is $0$ on the left, $0.98$ on the
right. The inclination is face-on in red, edge-on in green.
}
\label{fig:histoangleflux}
\end{figure*}
Fig.~\ref{fig:histoangleflux} shows the observed specific intensity distribution
as a function of the direction cosine $\mu$, for both spins. 
It shows that photons forming the face-on ray-traced spectrum are emitted
in a narrower range of values of $\mu$ than their edge-on counterparts.
This effect gets stronger with increasing spin: for $a=0.98$, photons forming
the edge-on ray-traced spectrum are coming from a very broad range of values
of $\mu$. This explains why edge-on ray-traced spectra are closer to the angle-averaged
solution, and even more at higher spins. The strong dependence on the spin parameter 
of this effect is explained
\begin{figure*}[htbp]
\centering
\includegraphics[width=16cm,height=7cm]{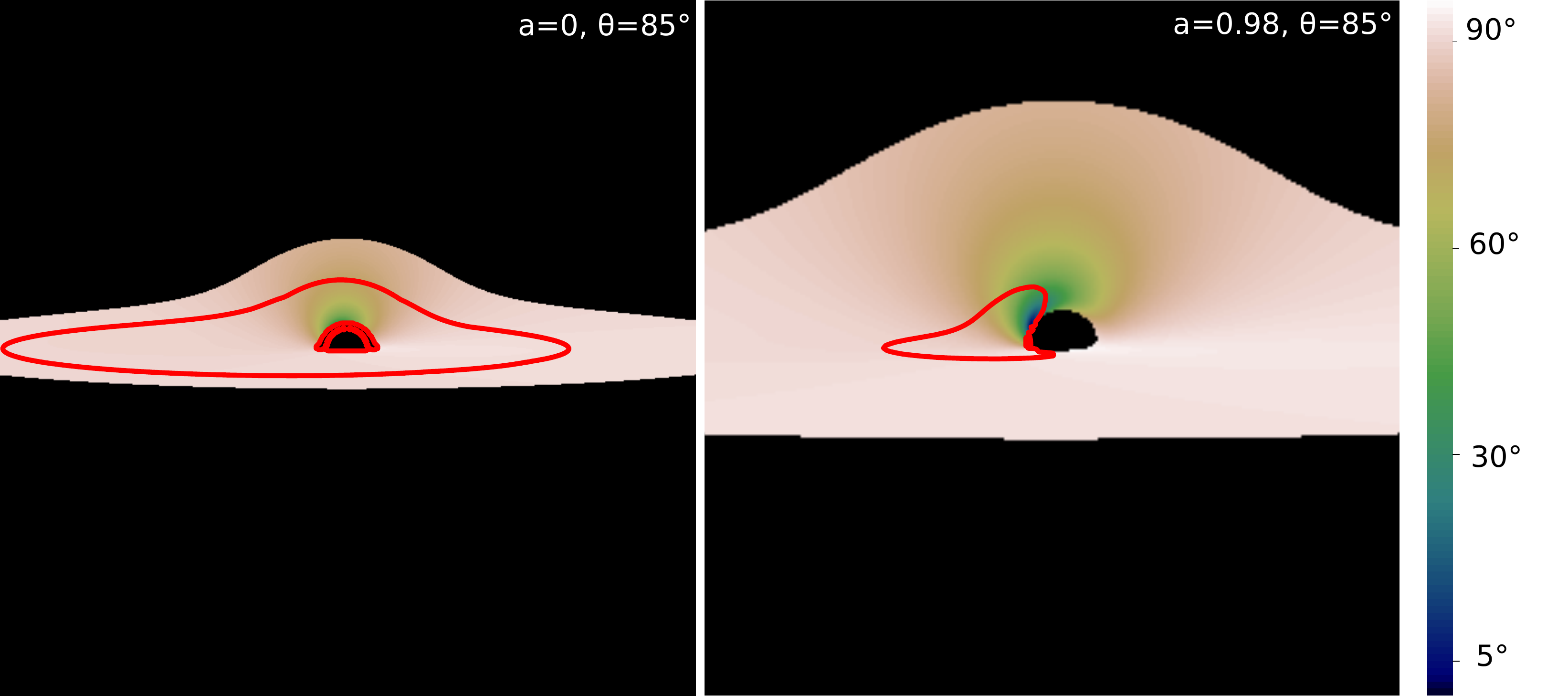}
\caption{{\it Angle map and intensity contour for edge-on inclination}.
The two panels show a map of the values of the emission angle $i$ for both spins at
edge-on inclination. Over-plotted is one specific intensity contour in red, encompassing
$90\%$ of the total specific intensity emitted at the disk's surface. The beaming
effect is already clear at spin zero: more intensity is emitted on the left part of the 
image where matter moves towards the observer. The beaming effect is extremely
strong at high spin, only a very small part of the disk emits most of the radiation observed.
}
\label{fig:mapanglecontourflux}
\end{figure*}
in Fig.~\ref{fig:mapanglecontourflux}. This Figure shows that the very strong relativistic
beaming effect at high spin has for consequence that most of the flux is emitted from
the very inner parts of the disk, where the emission angle $i$ varies a lot (see Fig.~\ref{fig:mapangle},
right panel). Thus, all possible values of emission angle contribute and the final
spectrum is closer to the averaged solution.

{As far as the emission-angle-dependence of the spectra are concerned, 
our results are in reasonable agreement with the findings of~\citet{garcia14}.
Our Fig.~\ref{fig:histoangleflux}, right panel, can be compared to Fig.~8, upper-right and lower-right panels
of~\citet{garcia14}, which are similar. This shows once again that the ray-tracing part
of the computation is very similar in both works. It is more complicated to go into a detailed
comparison of the observed spectra and their dependence on the emission angle $i$.
Our Fig.~\ref{fig:raytracedspectra} is different from the left panels of Fig.~8 of~\citet{garcia14},
and this is obviously due to the differences in the computation of local spectra. \citet{garcia14}
find a maximum difference between directional and angle-averaged spectra of $\gtrsim 20 \%$
(taking into account their results for $z=10\,M$ only of course), while we find a maximum
difference of up to $50 \%$. It is not possible to go further in the comparison without doing
a complete code comparison, which is not the aim of this paper. Still, the conclusions
of both completely independent treatments are reasonably similar, as far as the
angle dependence of the spectra is concerned.}

{The upper-right panel of Fig.~\ref{fig:raytracedspectra} shows no iron line for $a=0.98$, neither
in emission nor in absorption. This was already noticeable in the upper middle and right panels
of Fig.~\ref{fig:localspectraIllumDirec} showing the local spectra that exhibit very narrow absorption
lines due to the fact that the thermal bump covers the iron-line complex. The relativistic blurring of these
narrow lines has for consequence to completely remove the line from the observed spectrum. However,
the fact that the thermal bump covers the iron-line complex is due to our choice of parameters for which
the illumination luminosity is dominated by the accretion-induced luminosity. We have checked that for a much 
smaller accretion rate ($\dot{m} = 10^{-5}\,m_{\mathrm{Edd}}$ ; with the same illumination) the local spectrum at spin $0.98$
has a less prominent thermal bump leading to a clear iron line in emission, similarly to the $a=0$
case depicted in Fig.~\ref{fig:localspectraIllumDirec}. This fact shows that a detailed study of the
observed iron-line complex as a function of the ratio of the illuminating to accretion-induced luminosities
would be very interesting. Such an analysis goes beyond the scope of the present paper and we will
present it in a future article.}

The conclusion of this Section is that it is very important to take into account the
directionality of the emitted, local spectrum in order to predict the iron-line complex
of the observed spectrum. Indeed, the error on the observed spectrum caused by
angle averaging is as high as $50\%$ in the iron line region.

%\textbf{I should add also a comparison with the results of Javier et al. TBD.}

%At first sight, this result might seem surprising as we saw in Fig.~\ref{fig:mapangle}
%that essentially all values of angles are represented for an edge-on viewed
%disk due to strong bending of light rays. It thus might seem that the ray-traced 
%and local spectra should behave very differently. However, this is not so because
%at high inclination (edge-on), relativistic beaming is strong which translates
%in a small portion of the disk emitting most of the radiation (the part of the disk 
%moving in the direction of the observer). As a consequence for both face-on
%and edge-on viewed disk, only a small set of emission angles is probed.
%This is illustrated in Fig.~\ref{}

%\subsubsection{Impact of the thermal bump on the iron-line complex}

%\textbf{Andrzej, Agata, if you want to edit this part you are welcome :)}

\section{Conclusion}
\label{sec:conclu}

{The aim of this article is to compute X-ray spectra reprocessed by the accretion disk
of an X-ray binary in a very realistic way. We consider the isotropic emission of X-rays from
a nearly point-like lamp at rest at $z=10\,M$ along the black hole axis. We have computed
in full general relativity the irradiation onto the accretion disk for two extreme values of the spin
parameter $a=0$ and $a=0.98$. We then compute the local spectra reprocessed by the
accretion disk, taking self-consistently into account the black spin by solving the hydrostatic
equilibrium of the disk together with the radiative transfer. Finally we ray trace this local spectrum
to a distant observer in order predict a very realistic observed spectrum.}

{We show that taking into account the angle dependence of the local spectra is very
important for obtaining an accurate observed spectrum. In particular, the influence of
averaging out the angular information of the local spectra leads to an error in the
observed spectrum of $50 \%$ in the iron-line region, which is the important region 
to constrain black hole spins.}

\red{We also show that, for the set of parameters chosen in this work, the 
disk thermal emission is strong enough to cover the iron line complex
for spin $a=0.98$, resulting in
a featureless reflected spectrum.
On the other hand, the
direct component is much higher than the reflected one at a=0. Thus,
for both spins, the observed (direct+thermal+reflected) spectrum contains only weak reflection
features.}

\red{We intend to devote future work to investigating in detail the appearance of
the observed spectrum in various X-ray binary states (thermal dominated and hard state)
for a set of different lamp altitudes (which has a strong impact on the reprocessed/direct
components ratio). This will allow us to determine what sets of parameters allow to generate
reflection-dominated spectra for different source spectral states.}

\section*{Acknowledgements}
{
FHV, AR and AAZ acknowledge support 
from the National Science Center, Poland, under
grants: 2013/09/B/ST9/00060, 2013/11/B/ST9/04528, 2015/17/B/ST9/03422,
2012/04/M/ST9/00780 and 2013/10/M/ST9/00729. AR acknowledges support
from the Polish Ministry of Science and Higher Education grant W30/7.PR/2013.
This research was conducted within the scope of the HECOLS International Associated
Laboratory, supported in part by the National Science Center, Poland, grant
DEC-2013/08/M/ST9/00664. This research has 
received funding from the European Union Seventh Framework Program (FP7/2007-2013) 
under grant agreement No.312789.
}
%---------------------------------------------------------------------
%---------------------------------------------------------------------
%
%
%---------------------------------------------------------------------
%---------------------------------------------------------------------
\bibliography{ReflectedSpectra}

\begin{thebibliography}{48}
\expandafter\ifx\csname natexlab\endcsname\relax\def\natexlab#1{#1}\fi

\bibitem[{{Ballantyne} {et~al.}(2001){Ballantyne}, {Ross}, \&
  {Fabian}}]{ballantyne01}
{Ballantyne}, D.~R., {Ross}, R.~R., \& {Fabian}, A.~C. 2001, \mnras, 327, 10

\bibitem[{{Czerny} \& {Zycki}(1994)}]{czerny94}
{Czerny}, B. \& {Zycki}, P.~T. 1994, \apjl, 431, L5

\bibitem[{{Dauser} {et~al.}(2013){Dauser}, {Garcia}, {Wilms}, {B{\"o}ck},
  {Brenneman}, {Falanga}, {Fukumura}, \& {Reynolds}}]{dauser13}
{Dauser}, T., {Garcia}, J., {Wilms}, J., {et~al.} 2013, \mnras, 430, 1694

\bibitem[{{Done} {et~al.}(2007){Done}, {Gierli{\'n}ski}, \& {Kubota}}]{done07}
{Done}, C., {Gierli{\'n}ski}, M., \& {Kubota}, A. 2007, \aapr, 15, 1

\bibitem[{{Dovciak} \& {Done}(2015)}]{dovciak15}
{Dovciak}, M. \& {Done}, C. 2015, in The Extremes of Black Hole Accretion, 26

\bibitem[{{Dovciak} {et~al.}(2013){Dovciak}, {Matt}, {Bianchi}, {Boller},
  {Brenneman}, {Bursa}, {D'Ai}, {di Salvo}, {de Marco}, {Goosmann}, {Karas},
  {Iwasawa}, {Kara}, {Miller}, {Miniutti}, {Papadakis}, {Petrucci}, {Ponti},
  {Porquet}, {Reynolds}, {Risaliti}, {Rozanska}, {Zampieri}, {Zezas}, \&
  {Young}}]{dovciak13}
{Dovciak}, M., {Matt}, G., {Bianchi}, S., {et~al.} 2013, ArXiv e-prints

\bibitem[{{Dovciak} {et~al.}(2014){Dovciak}, {Svoboda}, {Goosmann}, {Karas},
  {Matt}, \& {Sochora}}]{dovciak14}
{Dovciak}, M., {Svoboda}, J., {Goosmann}, R.~W., {et~al.} 2014, ArXiv e-prints

\bibitem[{{Fabian} {et~al.}(1989){Fabian}, {Rees}, {Stella}, \&
  {White}}]{fabian89}
{Fabian}, A.~C., {Rees}, M.~J., {Stella}, L., \& {White}, N.~E. 1989, \mnras,
  238, 729

\bibitem[{{Fukumura} \& {Kazanas}(2007)}]{fukumura07}
{Fukumura}, K. \& {Kazanas}, D. 2007, \apj, 664, 14

\bibitem[{{Garc{\'{\i}}a} {et~al.}(2014){Garc{\'{\i}}a}, {Dauser}, {Lohfink},
  {Kallman}, {Steiner}, {McClintock}, {Brenneman}, {Wilms}, {Eikmann},
  {Reynolds}, \& {Tombesi}}]{garcia14}
{Garc{\'{\i}}a}, J., {Dauser}, T., {Lohfink}, A., {et~al.} 2014, \apj, 782, 76

\bibitem[{{Garc{\'{\i}}a} \& {Kallman}(2010)}]{garcia10}
{Garc{\'{\i}}a}, J. \& {Kallman}, T.~R. 2010, \apj, 718, 695

\bibitem[{{George} \& {Fabian}(1991)}]{george91}
{George}, I.~M. \& {Fabian}, A.~C. 1991, \mnras, 249, 352

\bibitem[{{Guilbert}(1981)}]{guilbert81}
{Guilbert}, P.~W. 1981, \mnras, 197, 451

\bibitem[{{Haardt} \& {Maraschi}(1991)}]{haardt91}
{Haardt}, F. \& {Maraschi}, L. 1991, \apjl, 380, L51

\bibitem[{{Kershaw}(1987)}]{kershaw87}
{Kershaw}, D.~S. 1987, \jqsrt, 38, 347

\bibitem[{{Madej}(1989)}]{madej89}
{Madej}, J. 1989, \apj, 339, 386

\bibitem[{{Madej}(1991)}]{madej91}
{Madej}, J. 1991, \apj, 376, 161

\bibitem[{{Madej} \& {R{\'o}{\.z}a{\'n}ska}(2000)}]{madej2000}
{Madej}, J. \& {R{\'o}{\.z}a{\'n}ska}, A. 2000, \aap, 363, 1055

\bibitem[{{Madej} \& {R{\'o}{\.z}a{\'n}ska}(2004)}]{madej04}
{Madej}, J. \& {R{\'o}{\.z}a{\'n}ska}, A. 2004, \mnras, 347, 1266

\bibitem[{{Madej} {et~al.}(2016){Madej}, {R{\'o}{\.z}a{\'n}ska}, {Majczyna}, \&
  {Nale{\.z}yty}}]{madej16}
{Madej}, J., {R{\'o}{\.z}a{\'n}ska}, A., {Majczyna}, A., \& {Nale{\.z}yty}, M.
  2016, ArXiv e-prints 1602.05088

\bibitem[{{Magdziarz} \& {Zdziarski}(1995)}]{magdziarz95}
{Magdziarz}, P. \& {Zdziarski}, A.~A. 1995, \mnras, 273, 837

\bibitem[{{Markoff} {et~al.}(2005){Markoff}, {Nowak}, \& {Wilms}}]{markoff05}
{Markoff}, S., {Nowak}, M.~A., \& {Wilms}, J. 2005, \apj, 635, 1203

\bibitem[{{Martocchia} \& {Matt}(1996)}]{martocchia96}
{Martocchia}, A. \& {Matt}, G. 1996, \mnras, 282, L53

\bibitem[{{Matt} {et~al.}(1991){Matt}, {Perola}, \& {Piro}}]{matt91}
{Matt}, G., {Perola}, G.~C., \& {Piro}, L. 1991, \aap, 247, 25

\bibitem[{{Miller} {et~al.}(2014){Miller}, {Mineshige}, {Kubota}, {Yamada},
  {Aharonian}, {Done}, {Kawai}, {Hayashida}, {Reis}, {Mizuno}, {Noda}, {Ueda},
  {Shidatsu}, \& {for the ASTRO-H Science Working Group}}]{miller14}
{Miller}, J.~M., {Mineshige}, S., {Kubota}, A., {et~al.} 2014, ArXiv e-prints
  1412.1173

\bibitem[{{Miniutti} \& {Fabian}(2004)}]{miniutti04}
{Miniutti}, G. \& {Fabian}, A.~C. 2004, \mnras, 349, 1435

\bibitem[{{Miniutti} {et~al.}(2003){Miniutti}, {Fabian}, {Goyder}, \&
  {Lasenby}}]{miniutti03}
{Miniutti}, G., {Fabian}, A.~C., {Goyder}, R., \& {Lasenby}, A.~N. 2003,
  \mnras, 344, L22

\bibitem[{{Nandra} {et~al.}(2013){Nandra}, {Barret}, {Barcons}, {Fabian}, {den
  Herder}, {Piro}, {Watson}, {Adami}, {Aird}, {Afonso}, \& et~al.}]{nandra13}
{Nandra}, K., {Barret}, D., {Barcons}, X., {et~al.} 2013, ArXiv e-prints

\bibitem[{{Nayakshin} \& {Kallman}(2001)}]{nayakshin01}
{Nayakshin}, S. \& {Kallman}, T.~R. 2001, \apj, 546, 406

\bibitem[{{Nied{\'z}wiecki} {et~al.}(2016){Nied{\'z}wiecki}, {Zdziarski}, \&
  {Szanecki}}]{niedzwiecki16}
{Nied{\'z}wiecki}, A., {Zdziarski}, A.~A., \& {Szanecki}, M. 2016, ApJL, in
  press, ArXiv e-prints 1602.09075

\bibitem[{{Nied{\'z}wiecki} \& {{\.Z}ycki}(2008)}]{niedzwiecki08}
{Nied{\'z}wiecki}, A. \& {{\.Z}ycki}, P.~T. 2008, \mnras, 386, 759

\bibitem[{{Pomraning}(1973)}]{pomraning73}
{Pomraning}, G.~C. 1973, {The equations of radiation hydrodynamics}

\bibitem[{{Poutanen} {et~al.}(1996){Poutanen}, {Nagendra}, \&
  {Svensson}}]{poutanen96}
{Poutanen}, J., {Nagendra}, K.~N., \& {Svensson}, R. 1996, \mnras, 283, 892

\bibitem[{{Reynolds} {et~al.}(2014){Reynolds}, {Ueda}, {Awaki}, {Gallo},
  {Gandhi}, {Haba}, {Kawamuro}, {LaMassa}, {Lohfink}, {Ricci}, {Tazaki},
  {Zoghbi}, \& {on behalf of the ASTRO-H Science Working Group}}]{reynolds14b}
{Reynolds}, C., {Ueda}, Y., {Awaki}, H., {et~al.} 2014, ArXiv e-prints
  1412.1177

\bibitem[{{Reynolds}(2014)}]{reynolds14}
{Reynolds}, C.~S. 2014, \ssr, 183, 277

\bibitem[{{Reynolds} \& {Nowak}(2003)}]{reynolds03}
{Reynolds}, C.~S. \& {Nowak}, M.~A. 2003, \physrep, 377, 389

\bibitem[{{Reynolds} {et~al.}(1999){Reynolds}, {Young}, {Begelman}, \&
  {Fabian}}]{reynolds99}
{Reynolds}, C.~S., {Young}, A.~J., {Begelman}, M.~C., \& {Fabian}, A.~C. 1999,
  \apj, 514, 164

\bibitem[{{Ross} \& {Fabian}(1993)}]{ross93}
{Ross}, R.~R. \& {Fabian}, A.~C. 1993, \mnras, 261, 74

\bibitem[{{R{\'o}{\.z}a{\'n}ska} {et~al.}(2002){R{\'o}{\.z}a{\'n}ska},
  {Dumont}, {Czerny}, \& {Collin}}]{rozanska02}
{R{\'o}{\.z}a{\'n}ska}, A., {Dumont}, A.-M., {Czerny}, B., \& {Collin}, S.
  2002, \mnras, 332, 799

\bibitem[{{R{\'o}{\.z}a{\'n}ska} \& {Madej}(2008)}]{rozanska08}
{R{\'o}{\.z}a{\'n}ska}, A. \& {Madej}, J. 2008, \mnras, 386, 1872

\bibitem[{{R{\'o}{\.z}a{\'n}ska} {et~al.}(2011){R{\'o}{\.z}a{\'n}ska}, {Madej},
  {Konorski}, \& {S{\c a}dowski}}]{rozanska11}
{R{\'o}{\.z}a{\'n}ska}, A., {Madej}, J., {Konorski}, P., \& {S{\c a}dowski}, A.
  2011, \aap, 527, A47

\bibitem[{{Ruszkowski}(2000)}]{ruszowski00}
{Ruszkowski}, M. 2000, \mnras, 315, 1

\bibitem[{{S{\c a}dowski} {et~al.}(2011){S{\c a}dowski}, {Abramowicz}, {Bursa},
  {Klu{\'z}niak}, {Lasota}, \& {R{\'o}{\.z}a{\'n}ska}}]{sadowski11}
{S{\c a}dowski}, A., {Abramowicz}, M., {Bursa}, M., {et~al.} 2011, \aap, 527,
  A17

\bibitem[{{Takahashi} {et~al.}(2014){Takahashi}, {Mitsuda}, {Kelley}, {Fabian},
  {Mushotzky}, {Ohashi}, {Petre}, \& {on behalf of the ASTRO-H Science Working
  Group}}]{takahashi14}
{Takahashi}, T., {Mitsuda}, K., {Kelley}, R., {et~al.} 2014, ArXiv e-prints
  1412.2351

\bibitem[{{Vincent} {et~al.}(2011){Vincent}, {Paumard}, {Gourgoulhon}, \&
  {Perrin}}]{vincent11}
{Vincent}, F.~H., {Paumard}, T., {Gourgoulhon}, E., \& {Perrin}, G. 2011,
  Classical and Quantum Gravity, 28, 225011

\bibitem[{{Wilkins} \& {Fabian}(2012)}]{wilkins12}
{Wilkins}, D.~R. \& {Fabian}, A.~C. 2012, \mnras, 424, 1284

\bibitem[{{Zdziarski} \& {Gierli{\'n}ski}(2004)}]{zdziarski04}
{Zdziarski}, A.~A. \& {Gierli{\'n}ski}, M. 2004, Progress of Theoretical
  Physics Supplement, 155, 99

\bibitem[{{Zycki} \& {Czerny}(1994)}]{zycki94}
{Zycki}, P.~T. \& {Czerny}, B. 1994, \mnras, 266, 653

\end{thebibliography}
\bibliographystyle{aa}
%---------------------------------------------------------------------
%---------------------------------------------------------------------
\end{document}